\documentclass[%
 reprint,
superscriptaddress,
 amsmath,amssymb,
 aps,
floatfix,
]{revtex4-2}

\usepackage{graphicx}
\usepackage{dcolumn}
\usepackage{bm}
\usepackage{hyperref}
\usepackage{float}
\usepackage[version=3]{mhchem}
\usepackage{comment}
\usepackage{layouts}

\usepackage{xcolor}
\usepackage{ulem}

\def\*#1{\mathbf{#1}}

\begin{document}



\title{Electronic localization and optical activity of strain-engineered transition-metal dichalcogenide nanobubbles}

\author{Stefan Velja}
 \affiliation{Institute of Physics, Carl von Ossietzky Universit\"at Oldenburg, 26129 Oldenburg, Germany}
 \affiliation{Friedrich-Schiller Universit\"at Jena, Institute for Condensed Matter Theory and Optics, 07743 Jena, Germany}
 
\author{Alexander Steinhoff}
 \affiliation{Institute of Physics, Carl von Ossietzky Universit\"at Oldenburg, 26129 Oldenburg, Germany}

\author{Jannis Krumland}
 \affiliation{Institute of Physics, Carl von Ossietzky Universit\"at Oldenburg, 26129 Oldenburg, Germany}

 \author{Christopher Gies}
 \affiliation{Institute of Physics, Carl von Ossietzky Universit\"at Oldenburg, 26129 Oldenburg, Germany}

\author{Caterina Cocchi}
 \affiliation{Institute of Physics, Carl von Ossietzky Universit\"at Oldenburg, 26129 Oldenburg, Germany}
 \affiliation{Center for Nanoscale Dynamics (CeNaD), Carl von Ossietzky Universit\"at Oldenburg, 26129 Oldenburg, Germany}
 \affiliation{Friedrich-Schiller Universit\"at Jena, Institute for Condensed Matter Theory and Optics, 07743 Jena, Germany}

\date{\today}

\begin{abstract}
Strain-engineered transition-metal dichalcogenide nanobubbles are promising platforms for quantum emission, as revealed by recent experimental observations. In this work, we present an \textit{ab initio} investigation of MoS$_2$, WS$_2$, MoSe$_2$, and WSe$_2$ nanobubbles, linking their structural and electronic properties to predictions of their optical activity. Inflating forces yield tunable geometries with non-uniform, apex-concentrated strain, which is sensitive to material rigidity. Strain modifies band gaps and universally induces non-dispersive valence states, exhibiting composition-dependent wave-function character, as revealed by an in-depth analysis of band structures and orbital contributions. Crucially, transitions from these apex-localized valence states are predominantly dark. This characteristic is attributed to their localization at the $\Gamma$-point, inhibiting transitions to the lowest unoccupied states that reside at the K-valley. While revealing that the herein considered sub-10-nm nanobubbles fall short as single-photon emitters, our findings provide essential understanding of the structure-property relations in emerging quantum materials, providing robust design rules to optimize their characteristics for novel quantum applications.
\end{abstract}

\maketitle

\section{Introduction}
The advent of the second quantum revolution demands suitable material platforms that combine favorable electronic and optical properties with flexible and compatible structures for device integration~\cite{levy22aplm}. Since their discovery~\cite{cole+11sci}, transition-metal dichalcogenide (TMDC) monolayers have revealed themselves as ideal quantum materials~\cite{liu-hers19nrm}. Their tunable band gaps and valley-rich band structure, their pronounced excitonic features, and favorable characteristics for fabrication, heterostructuring, and device integration have kept them in the spotlight for over a decade~\cite{fior+14natn,wang+18rmp,avsa+20rmp}. Their structural flexibility, in particular, enables strain and mechanical deformations to act as a non-invasive knob for tailoring physical properties, including band-gap size and exciton binding energies~\cite{duer+14natcom,song+16nl,mait+20natph}. 

Recent experimental findings have shown that TMDC-based nanoprotrusions can be created by inflating monolayers with gas molecules~\cite{darlington2020imaging,blun+20prr,blun+21prl,stel+24nl} or through deposition on a nanopatterned substrate~\cite{part+21natcom,kim+22nano,li+22nano,shab+22nl}. The resulting nanostructures -- often called ``nanobubbles'' -- can host spatially and energetically localized electronic states, which are crucial for quantum emission~\cite{azza+21apl}. Most reports on  quantum emission in strained TMDCs have been linked to the presence of defects~\cite{,part+21natcom,yu+25nl,blun+25nl} or achieved on larger strain features~\cite{darlington2020imaging,blun+21prl,kim+22nano}, where the localized light emission was clearly identified. While exciton localization has been reported in nanobubbles~\cite{darlington2020imaging,shab+22nl,chen+22ami}, direct proof of single-photon emission purely due to strain-induced quantum confinement in sub-10-nm nanobubbles remains an experimental challenge.
The electronic and optical properties of TMDC nanobubbles have been rationalized in several theoretical studies~\cite{darlington2020imaging,gast+23npj2dma,stei+25prb}, even combining semi-classical atomistic methods and quantum models~\cite{carm+19nl}. However, this knowledge, although often obtained in tandem with measurements~\cite{darlington2020imaging,stei+25prb}, is insufficient to predict how controlled strain and chemical composition systematically influence the structural, electronic, and optical characteristics of pristine TMDC nanobubbles. Gaining insight independently of experiments is essential to isolate and establish the fundamental limits of quantum confinement induced purely by strain, particularly given the experimental challenges of precisely controlling atomistic degrees of freedom, such as defects and substitutions. 

\textit{Ab initio} quantum-mechanical methods such as density-functional theory (DFT) are ideally suited for this purpose, enabling the simulation of (in principle) any compound given its atomic species and mutual arrangement. While currently affordable computational efforts typically limit these simulations to the 10-nm size, this regime is of particular physical interest as it maximizes the effects of curvature and quantum confinement.
The importance of these studies is additionally underscored by the current fabrication and characterization challenges related to such 10-nm structures~\cite{chen+25nl}, making computational experiments a primary tool for exploring and rationalizing this nanoscale manipulation.
For example, in recent DFT work, some of us characterized the electronic structure of one-dimensional \ce{MoSe2} nanowrinkles~\cite{velj+24ns}. Furthermore, we revealed the presence of a non-dispersive mid-gap state in the electronic structure of sub-10-nm \ce{MoS2} nanobubbles, whose energy and multiplicity can be further modulated by sulphur single vacancies~\cite{krumland2024quantum}. Clarifying whether this localized state is a universal characteristic of TMDC nanobubbles of similar size but varying composition (i.e., by replacing Mo with W and/or S with Se), assessing its energy as a function of strain, and evaluating the optical activity of the pristine lattice is essential information to optimize these nanostructures for quantum applications. 

In this work, we investigate the structural and electronic properties of 36 defect-free, free-standing TMDC nanobubbles with chemical formula \ce{MX2} (M = Mo, W and X = S, Se), subject to 9 inflating forces ranging from zero (flat monolayers) to 0.02~a.u./atom (atomic units per atom). Using DFT, we optimize their structures and provide insight into their characteristics, including the relationship between strain and morphology. We then proceed with the analysis of the electronic properties, focusing on the energy and localization of the non-dispersive state in the valence band, a common feature of all considered nanobubbles. With this information, we finally evaluate the optical activity of the transition between this state, especially when it appears in the mid-gap, and the lowest unoccupied band, providing critical insights into the viability of these nanostructures as single-photon emitters based purely on strain and quantum confinement.

\section{Methodology}

All calculations presented in this work are performed in the framework of density-functional theory~\cite{hohenberg1964inhomogeneous}, adopting the Kohn-Sham (KS) scheme~\cite{kohn1965self} as implemented in the pseudopotential plane-wave code \texttt{Quantum ESPRESSO}~\cite{giannozzi2017advanced}. The KS equation 
\begin{equation}
    \left(-\frac{\hbar^2}{2m_\mathrm{e}}\nabla^2 + v_{\mathrm{eff}} -\varepsilon_{n\*k}\right) \psi_{n\*k}(\*r) = 0
\label{eq:KS}
\end{equation}
yields electron energies $\varepsilon_{n\*k}$ and wave-functions $\psi_{n\*k}(\*r)$ upon diagonalization of the single-particle KS Hamiltonian, which includes the kinetic energy term $(-\frac{\hbar^2}{2m_\mathrm{e}}\nabla^2)$, where $m_e$ is the electron mass, and an effective potential $v_\mathrm{eff}$. In turn, $v_\mathrm{eff}$ is the sum of three terms: the electron-nuclear interaction, the Hartree potential, and the exchange-correlation (XC) potential. The latter is herein approximated by the Perdew-Burke-Ernzerhof (PBE)~\cite{perdew1996generalized} version of the generalized-gradient approximation. 
Core electrons of all elements are described by SG15 optimized norm-conserving Vanderbilt pseudopotentials~\cite{schlipf2015optimization}. For projected density of states calculations, ultrasoft pseudopotentials \cite{bennett2019systematic} are adopted for Se, S, and W atoms. Spin-orbit coupling (SOC) is included in all calculations except for structural optimization, where we checked that excluding this effect does not alter the results while considerably speeding up the calculations. A convergence threshold of $10^{-6}$~Ry is set for both relaxation and self-consistent runs. Interatomic forces are minimized until they are below 50~meV/\AA{}. Kinetic energy cutoffs for wave functions and charge density are set to 816~eV and 3265~eV, respectively. 

A vacuum layer of 20~\AA{} is applied in the out-of-plane direction to decouple each supercell from its replicas. We checked against larger vacuum thicknesses that this setup is sufficient to completely decouple periodic replicas in the out-of-plane direction. Band-structure plots are obtained using \texttt{Wannier90}~\cite{pizzi2020wannier90} at the $\Gamma$-point only, taking advantage of the very small size of the Brillouin zone of the considered systems and the fact that primitive-cell $\*k$ is no longer a good quantum number due to the breaking of translational symmetry. Other electronic and optical properties are likewise calculated at the $\Gamma$-point only, using their respective tools.

Optical transition tensor elements 
\begin{equation}
    \hat{\*M}_{\alpha,\beta} = \langle u_{\*{k},n'} | \hat{\*p}_\alpha | u_{\*{k},n} \rangle \langle u_{\*{k},n} | \hat{\*p}_\beta^\dagger | u_{\*{k},n'} \rangle,
    \label{eq:opticaltensor}
\end{equation}
where $u$ is the periodic part of the Bloch function, $\hat{\*p}_\alpha$ and $\hat{\*p}_\beta$ are the components of the momentum operator, are computed in the independent particle approximation using Fermi's golden rule, as implemented in the \texttt{Epsilon.x} routine of \texttt{Quantum ESPRESSO}.



\section{Results and discussion}

\subsection{Structural properties}
\begin{figure}
    \centering
    \includegraphics[width=0.48\textwidth]{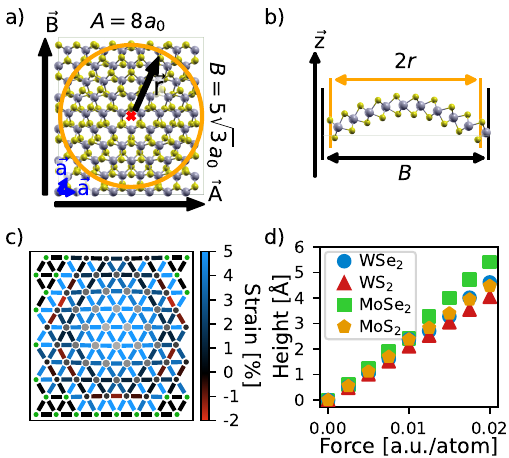}
        \caption{a) Ball-and-stick representation, produced by the visualization software \texttt{XCrySDen}~\cite{kokalj1999xcrysden}, of the TMDC nanobubble supercell, with metal atoms in gray and chalcogen atoms in yellow. The lower-layer chalcogen atoms inside the orange circle with radius $r$ experience the inflating force. The metal atoms outside of the circle are held fixed. All other atoms are free to relax. b) Nanobubble profile, with orange bars defining the region of applied strain, while the black bars mark the supercell boundaries.
    c) Local strain map of a \ce{MoSe2} nanobubble subject to $F = 0.0175$~a.u./atom given by the relative change in interatomic distance between neighbouring metal atoms. Red and blue segments represent regions of compressive and tensile local strain, respectively, with values capped at +5\% for visualization. Gray dots indicate metal atoms under the inflating force, with lighter shades indicating larger elevation. Green dots mark the positions of the metal atoms held fixed during relaxation. d) Maximum height of the nanobubbles as a function of the inflating force.}
    \label{fig:structurePanels}
\end{figure}

We begin with the analysis of the structural properties of the considered TMDC nanobubbles, modeled in orthorhombic supercells derived from the hexagonal primitive unit cells with lattice parameters $a_0$ optimized in this work and in agreement with the literature \cite{rama12prb, haastrup2018computational, gjerding2021recent}, see \autoref{tab:struct}. The supercell lattice constants are $A = 8a_0$ and $B = 5\sqrt{3}a_0$ (see
\autoref{fig:structurePanels}a and \autoref{tab:struct}). Due to periodic boundary conditions, an in-plane array of nanobubbles is formed (see Figure~S1 of the Supplemental Material). 

The nanobubbles are generated by applying an inflating force $F$, to the bottom-layer chalcogen atoms within the circular region with radius $r$ visualized in \autoref{fig:structurePanels}a. The central strained area contains 56 metal atoms and 53 chalcogen atoms in each chalcogen layer. The applied force, ranging from 0.00 to 0.02~a.u./atom with equispaced increments of 0.0025~a.u./atom, further displaces the central metal and upper-layer chalcogen atoms from their equilibrium positions, forming a nanobubble with broken mirror symmetry across the metal layer. Each inflated nanostructure is obtained by relaxing a pristine monolayer under the applied forces. During this optimization, all atoms are allowed to move freely, except for metal atoms outside the circular region with radius $r$, which are held fixed (Figure~S2 in the Supporting Information), mimicking the pinning of the layer to the substrate through Van der Waals forces. The nanobubble profile is shown in \autoref{fig:structurePanels}b.

\begin{table}[h!]
\caption{Structural parameters (in \AA{}) of the TMDC nanobubbles: in-plane lattice vector $a_0$ of the primitive hexagonal cell optimized in this work, lattice parameters of the orthorhombic supercell, $A$ and $B$, and radius $r$ defining the area in the supercells where the atoms are subject to inflation.}
\label{tab:struct}
    \centering
    \begin{tabular}{c||c|c|c|c}
    \hline \hline
        & $a_0$ & $A$ & $B$ & $r$ \\ \hline \hline
        \ce{MoS2} & 3.18 & 25.48 & 27.58 & 12.50 \\ \hline
        \ce{MoSe2} & 3.30 & 26.39 & 28.57 & 12.95 \\ \hline
        \ce{WS2} & 3.19 & 25.50 & 27.60 & 12.44 \\ \hline
        \ce{WSe2} & 3.32 & 26.56 & 28.75 & 13.00 \\ \hline \hline

    \end{tabular}
\end{table}

The resulting nanobubbles exhibit structural characteristics that vary with the applied force and chemical composition. Their height, defined as the $z$-coordinate of the uppermost metal atom (with the fixed metal atoms set at $z=0$), shows a linear dependence on the inflating force for all TMDCs (\autoref{fig:structurePanels}d). \ce{MoSe2} nanobubbles achieve the greatest heights under all forces with 5.4~\AA{} at $F=0.02$~a.u./atom. In \ce{WSe2}, the nanobubble height is generally reduced (e.g., 4.6~\AA{} at $F=0.02$~a.u./atom). Interestingly, \ce{MoS2} nanobubbles show a height-force trend very similar to \ce{WSe2}, with a comparable maximum height of 4.5~\AA{} under the largest deformation. \ce{WS2} nanobubbles consistently exhibit the lowest heights (the maximum is approximately 4~\AA{} for $F=0.02$~a.u./atom), indicating they are the stiffest among the studied compounds. We note that this observed ordering of effective deformation, with \ce{WS2} being the stiffest material and \ce{MoSe2} the most flexible one, departs from the trend predicted by their macroscopic Young's modulus~\cite{bertolazzi2011stretching, liu2014elastic, falin2021mechanical, zeng2015electronic, cooper2013nonlinear}. This deviation is attributed to finite-size effects and non-linear mechanics inherent to the sub-10-nm scale of the considered herein nanobubbles, where the structural response is governed by localized bond bending and boundary conditions, rather than the bulk continuum mechanical properties.

As the TMDC monolayers deform into nanobubbles, local domains of tensile and compressive strain appear. In this context, strain is defined as the relative change in distance between neighboring metal atoms compared to the corresponding flat monolayer. For instance, a \ce{MoSe2} nanobubble subject to $F = 0.0175$~a.u./atom (\autoref{fig:structurePanels}c) is dominated by tensile strain domains, particularly at the apex of the deformation in the middle of the supercell. A perimeter of compressive strain surrounds this central region, consistent with the curved profile. Notably, additional tensile strain appears near the supercell edges, connecting relaxed metal atoms with those held fixed. The local strain pattern presented for the \ce{MoSe2} nanobubble in \autoref{fig:structurePanels}c consistently appears in all studied systems. A more detailed analysis of strain distribution, calculated in terms of Gaussian curvature~\cite{ONeill2006}, is reported in the Supporting Information (see Figure~S3 and related discussion). Composition-dependent trends are presented in Figure~S4 and Table~S1.

\subsection{Electronic properties}
\label{section:elec_properties}

\begin{figure}[h]
    \centering
    \includegraphics[width=0.48\textwidth]{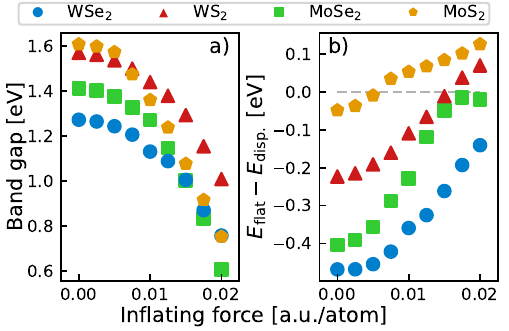}
    \caption{a) Band gap and b) energy difference between the flat state and the highest occupied dispersive state as a function of the inflating force. 
    }
    \label{fig:gapstates}
\end{figure}

We now examine the electronic properties of the considered nanobubbles starting from the fundamental band gaps as a function of the inflating force (\autoref{fig:gapstates}a). 
Under weak deformations ($F \leq 0.05$~a.u./atom), \ce{MoS2} nanobubbles show the largest band gaps, followed by \ce{WS2}, \ce{MoSe2}, and \ce{WSe2}. It is worth stressing that the difference from the usual ordering of the band gaps among these compounds \cite{manzeli20172d, zeng2015electronic, tongay2013defects, rama12prb, krustok2017local} stems from using the PBE functional in conjunction with SOC and is already present in literature \cite{rama12prb}. Since the SOC-induced splitting is approximately three times larger in the W-based compounds than in their Mo-based counterparts~\cite{kosmider2013large, manzeli20172d}, this causes a more prominent reduction in \ce{WS2} and \ce{WSe2} compared to their Mo-based siblings.
While all band gaps exhibit a parabolic decrease as a function of inflation, the trends vary significantly with the composition. For \ce{MoS2}, the band gap decreases steeply: At $F = 0.01$~a.u./atom, its gap is 1.35~eV, lower than that of \ce{WS2} (1.44~eV). By $F = 0.02$~a.u./atom, the \ce{MoS2} gap falls below 0.75~eV, comparable with \ce{WSe2} under the same force, and significantly smaller than the gap of \ce{MoSe2}, which reaches 0.6~eV. 
We recall that while the adopted PBE functional typically underestimates band gaps~\cite{crowley2016resolution, mori2008localization, borlido2020exchange}, a large error cancellation between quasi-particle corrections and exciton binding energies makes the PBE gaps in TMDCs within a range of 0.2-0.3~eV of the experimental optical gaps~\cite{qiu2013optical, komsa2012effects}. We anticipate similar trends in the nanobubbles and plan further investigation into this aspect in future work.

The varying band-gap trends are connected to the appearance of a non-dispersive (flat) state at the top of the valence band. This feature, previously discussed for \ce{MoS2} nanobubbles~\cite{krumland2024quantum}, appears as a universal feature across all nanostructures studied here. 
By inspecting \autoref{fig:gapstates}b, which displays the energy of the localized state relative to the valence-band maximum (VBM) as a function of the inflating force, we notice that in \ce{MoS2}, a force $F = 0.0075$~a.u./atom is sufficient to bring this localized state into the fundamental gap of the monolayer. Even above this threshold, its energy continues to increase with force. \ce{WS2} nanobubbles show similar behavior, although requiring a larger force ($F \geq 0.015$~a.u./atom) to push the flat state into the gap. In \ce{MoSe2} and \ce{WSe2} nanobubbles, on the other hand, the localized state remains within the valence band, see \autoref{fig:gapstates}b. In \ce{WSe2}, the applied force is insufficient to push the localized state high enough in energy to reach the VBM. In \ce{MoSe2}, the energy of this localized state steadily increases up to $F=0.0175$~a.u./atom, becoming nearly degenerate with the uppermost dispersive band but not crossing the VBM even upon the largest applied force $F=0.02$~a.u./atom (\autoref{fig:gapstates}b). Here, the dispersive valence bands remain so high in energy that the localized state cannot overcome them. 

\begin{figure*}
    \centering
    \includegraphics[width=0.98\textwidth]{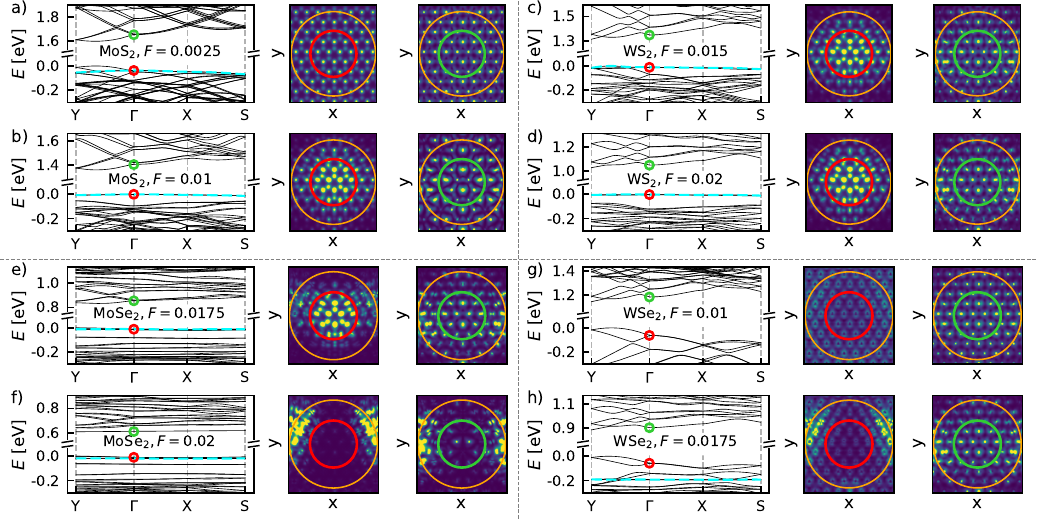}
    \caption{Band structures of selected TMDC nanobubbles with the flat states highlighted by cyan dashed lines. The VBM is set to 0.0~eV. The heat maps show the wave function distribution (square modulus) of the states marked by red and green dots in the band structures, with the red and green circles marking the apex region (20\% of the total supercell area) for the uppermost valence band and the lowest conduction band, respectively. The orange circles denote the area in which the force is applied. The composition and inflating force (in a.u./atom) of each nanobubble is reported as an inset in the corresponding band structure: a) and b) \ce{MoS2} at $F=0.0025$~a.u./atom and $F=0.01$~a.u./atom; c) and d) \ce{WS2} at $F=0.015$~a.u./atom and $F=0.02$~a.u./atom; e) and f) \ce{MoSe2} at $F=0.0175$~a.u./atom and $F=0.02$~a.u./atom; g) and h) \ce{WSe2} at $F=0.001$~a.u./atom and $F=0.0175$~a.u./atom.}
    \label{fig:BS-WFD-examples}
\end{figure*}

To further characterize the electronic properties of the TMDC nanobubbles, we examine the band structures along with the wave function distributions (WFDs) of selected frontier states at $\Gamma$ (\autoref{fig:BS-WFD-examples}), where the localized states at the nanobubble apex appear. This characteristic contrasts with the large dispersion of the electronic states of pristine TMDCs around the high-symmetry point K, where the most relevant optical and excitonic activity of these materials occurs. A detailed mapping between the state dispersion in the nanobubbles and in the unit cells of the TMDCs is reported below to support the discussion on optical transitions. Here, it is sufficient to focus on the behavior at the $\Gamma$-point to gain a deeper understanding of electronic state localization as a function of inhomogeneous strain.

For \ce{MoS2} at $F= 0.0025$~a.u./atom, the localized state is near but still below the VBM, with a spatial distribution spread across the supercell, similar to the conduction band minimum (CBm) at $\Gamma$ (\autoref{fig:BS-WFD-examples}a). At $F \leq 0.01$~a.u./atom, the localized state enters the fundamental gap, and its wave function is localized at the apex (\autoref{fig:BS-WFD-examples}b). The CBm at $\Gamma$ remains delocalized but its character slightly changes with increasing deformation. The behavior of the \ce{WS2} nanobubbles is qualitatively similar. At $F \leq 0.015$~a.u./atom, the non-dispersive occupied state at $\Gamma$ is localized at the apex, although the (dispersive) VBM is elsewhere in the Brillouin zone. At $F= 0.02$~a.u./atom, the flat band is in the mid-gap and its real-space distribution at $\Gamma$ becomes even more focused at the apex (\autoref{fig:BS-WFD-examples}d). In both \ce{WS2} nanobubbles, the lowest unoccupied state at $\Gamma$ remains spread across the entire supercell, although its localization toward the non-deformed edges increases with force. 

The \ce{MoSe2} nanobubbles at the two largest inflating forces, $F= 0.0175$~a.u./atom and $F= 0.02$~a.u./atom, offer an intriguing scenario. As shown in \autoref{fig:BS-WFD-examples}e,f, their uppermost valence bands do not form the dispersive manifold characteristic of the other compounds. At $F= 0.0175$~a.u./atom, two nearly degenerate states appear at the valence band top, one of which is localized at the apex (\autoref{fig:BS-WFD-examples}c). 
This departure from the general trend is not an artifact but a result of a near-degeneracy between the localized flat valence state and the highest occupied dispersive state, leading to strong state mixing and an abrupt, strain-induced electronic rearrangement. This behavior, which can be solved by slightly extending the nanobubble size with a single row of edge atoms not subject to strain (see Figure~S13), highlights the structural sensitivity and large electronic tunability of these systems. Additional results for \ce{WS2} nanobubbles with larger separations are reported in Figures~S10-S12.

At $F= 0.02$~a.u./atom, the valence band is not significantly altered, in contrast to the wave-function distribution, which is localized in the flat region at the supercell edges (\autoref{fig:BS-WFD-examples}f). The CBm at $\Gamma$ exhibits similar characteristics, being itself a flat state in the band structure, in contrast to its dispersive nature at $F= 0.0175$~a.u./atom (\autoref{fig:BS-WFD-examples}e). This is reflected in the flat CBm being spatially localized at the edges of the bubble. Finally, for the \ce{WSe2} nanobubble with $F= 0.01$~a.u./atom, both the VBM and CBm at $\Gamma$ are dispersive and delocalized, similar to the flat pristine monolayer (\autoref{fig:BS-WFD-examples}g). At $F= 0.0175$~a.u./atom, a flat state is visible in the band structure, remaining about 200~meV below the VBM at $\Gamma$, which is generally delocalized (\autoref{fig:BS-WFD-examples}h). While the CBM that is uniformly distributed in the supercell, the highest valence state at $\Gamma$ exhibits charge localization in the flat regions at the supercell edges.

\begin{figure}[h]
    \centering
    \includegraphics[width=0.48\textwidth]{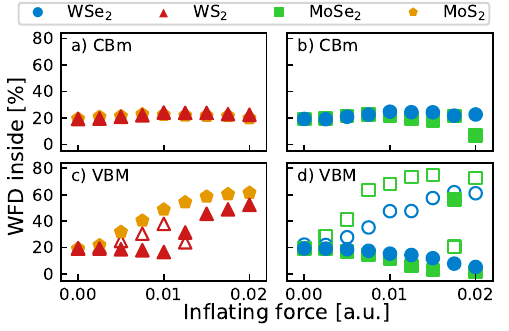}
    \caption{Wave-function distribution (WFD) at $\Gamma$ for the VBM and CBm localized at the apex encompassing 20\% of the total supercell area. Empty symbols indicate the localized states within the valence band or nearly degenerate with the VBM at $\Gamma$ if the VBM is a flat state. 
    }
    \label{fig:localization}
\end{figure}

The previous analysis indicates that the inflating forces affect the occupied electronic states of the TMDC nanobubbles, particularly by introducing a non-dispersive state localized at the apex, with increasing energy under increasing deformation. Since wave-function localization at the frontier is crucial for enabling single-photon emission~\cite{reed1988observation, tonndorf2015single, koperski2015single, linhart2019localized, moody2018microsecond, darlington2020imaging, srivastava2015optically, chakraborty2015voltage}, it is worth quantifying it by integrating the WFD of the flat states at $\Gamma$ within a circular region around the apex covering 20\% of the total supercell area (colored circles in the heatmaps of \autoref{fig:BS-WFD-examples}). As shown in \autoref{fig:localization}a,b (values reported in Tables~S1-S3), in all systems, the CBm localization at $\Gamma$ remains around 20\% as in the flat monolayers (\autoref{fig:localization}a,b). The only exception is \ce{MoSe2} at $F= 0.02$~a.u./atom, where the CBm corresponds to a localized state at the 
supercell edges, see \autoref{fig:BS-WFD-examples}f.

The VBM exhibits a more complex behavior. In the \ce{MoS2} nanobubbles, the VBM localization increases significantly with the inflating force (\autoref{fig:localization}c). For $F > 0.0075$~a.u./atom, the VBM coincides with the non-dispersive state localized at the apex (compare \autoref{fig:BS-WFD-examples}b), and its localization value is almost 50\% of the total density. While being reduced at weaker deformations, it further localizes for $F>0.01$~a.u./atom, reaching a plateau of approximately 61\% towards $F = 0.02$~a.u./atom (\autoref{fig:localization}c).
For \ce{WS2}, we notice a similar behavior under large inflating forces ($F \geq 0.015$~a.u./atom), where the VBM coincides with the localized state (compare \autoref{fig:gapstates}b). In this regime, the apex localization ranges from approximately 46\% at $F = 0.015$~a.u./atom to 52\% for $F = 0.02$~a.u./atom, see \autoref{fig:localization}c. For $F < 0.015$~a.u./atom, the VBM of the \ce{WS2} nanobubbles corresponds to a dispersive band, showing significant localization away from the apex, with values $\leq$ 20\%. Again, in these cases, the wave function tends to be more localized toward the flat supercell edges, as discussed above (see also Figs.~S2-S7). In the \ce{MoSe2} and \ce{WSe2} nanobubbles, where the flat state reaches or crosses the VBM, localization trends are more distinct. As shown in \autoref{fig:localization}d, the VBM localization (filled symbols) around the apex decreases monotonically with force, approaching zero for both compounds at $F = 0.02$~a.u./atom. This indicates that the VBM shifts its localization predominantly to the supercell edges, in the non-inflated regions between nanobubble replicas (see, e.g., \autoref{fig:BS-WFD-examples}h). Conversely, the localized valence states themselves (empty symbols) exhibit an almost monotonically increasing distribution around the apex, up to $\sim$75\% in the \ce{MoSe2} nanobubble with $F=0.015$~a.u./atom. A deeper analysis of the orbital character based on the Clebsch-Gordan coefficients is reported in the Supporting Information (Figure~S9 and related discussion).

\subsection{Optical activity}

\begin{figure*}
    \centering
    \includegraphics[width=0.98\textwidth]{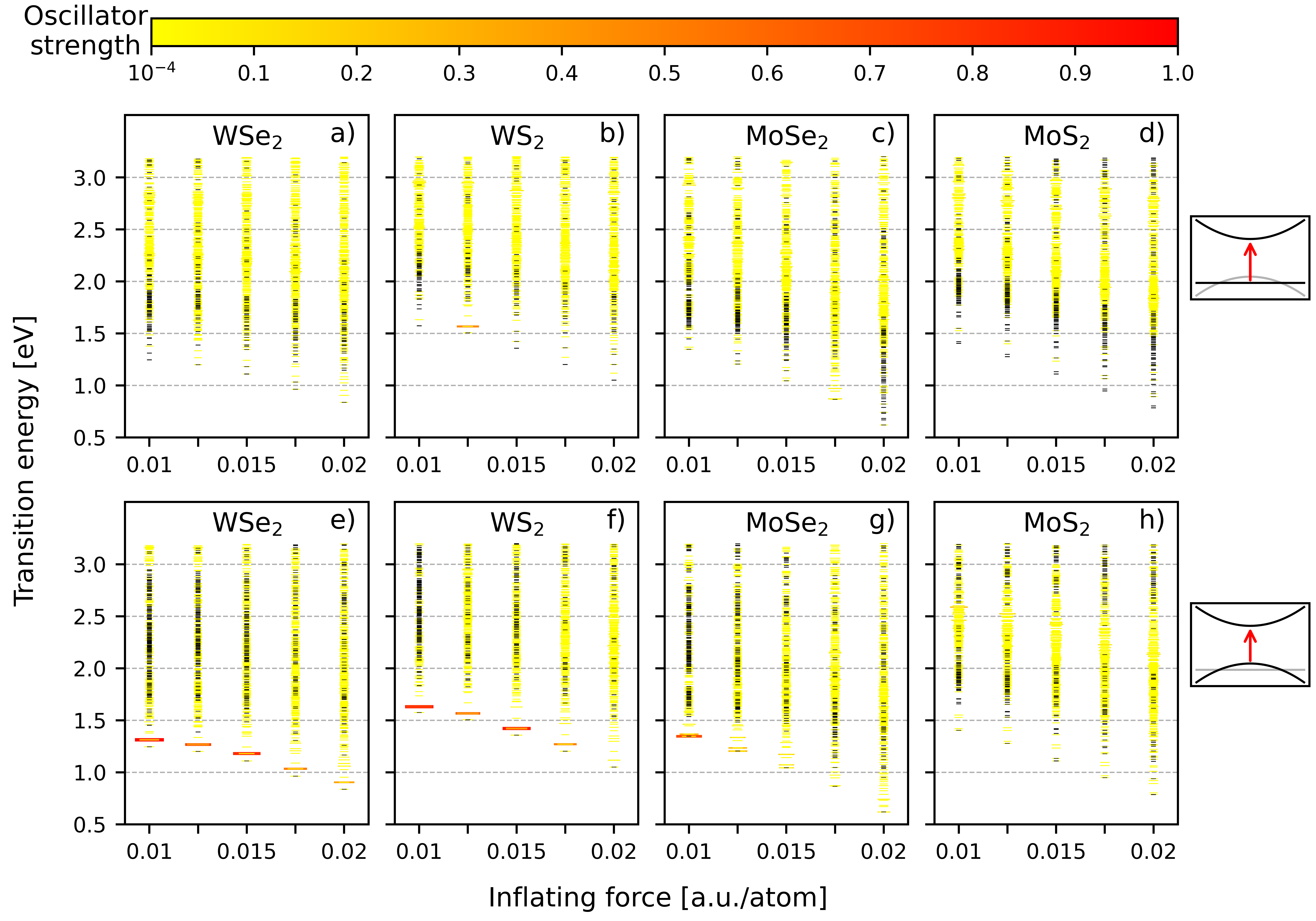}
    \caption{Transition energy, defined as the energy difference between the initial and final state, of the average in-plane optical matrix elements for  transitions at $\Gamma$ from the flat states of a) \ce{WSe2}, b) \ce{WS2}, c) \ce{MoSe2}, and d) \ce{MoS2} nanobubbles and from the uppermost occupied dispersive state of e) \ce{WSe2}, f) \ce{WS2}, g) \ce{MoSe2}, and h) \ce{MoS2} nanobubbles subject to a force $F \geq 0.01$~a.u./atom (see insets on the right). The oscillator strength is indicated by the color bar on top where transitions weaker than $<10^{-4}$ are considered dark and depicted in black. The width and thickness of the plotted bars increase (nonlinearly) with the transition energy, purely for visibility and clarity.
    }
    \label{fig:transition_matrix}
\end{figure*}

To complete our analysis, we evaluate the oscillator strength of optical transitions at $\Gamma$ from the top of the valence band to the lowest conduction band using Eq.~\eqref{eq:opticaltensor}. In doing so, we consider two scenarios: in the first one, the transition starts from the flat state; in the second one, the initial state is the highest dispersive band. In both cases, we focus only on the nanobubbles subjected to inflating forces equal to or larger than 0.01~a.u./atom, as they most significantly modify the electronic structure of the systems (\autoref{fig:gapstates}). As shown in \autoref{fig:transition_matrix}a-d, where the length and color of the bars (from short to long and from light to dark, with forbidden excitations in black) are representative of the oscillator strength, transitions from the non-dispersive uppermost valence band to the conduction states at $\Gamma$ are very weak if not completely dark. This trend is consistently preserved regardless of the nanobubble composition, with the transition to the CBm being forbidden or extremely weak in all considered nanobubbles. 

The only significant exception is found in \ce{WS2} subject to $F = 0.0125$~a.u./atom, which exhibits a strain-induced resonance between the localized apex state and the dispersive valence bands. At this specific inflating force, the energy of the localized state coincides with the uppermost dispersive band at the $\Gamma$-point (Figure~S8). This proximity leads to significant wave-function mixing between the dispersive bands and the localized state, which in turn enhances the oscillator strength of the corresponding optical transition. As the force increases further to $F = 0.015$ a.u./atom, the localized state is pushed higher into the gap, decoupling from the dispersive manifold and becoming dark again. A similar behavior is also found in \ce{MoSe2} under $F = 0.0175$~a.u./atom (Figure~\ref{fig:BS-WFD-examples}e). However, in that case, the increase in optical activity is noticeably weaker, due to the larger energetic distance from the dispersive state in this configuration.

An overall inspection of \autoref{fig:transition_matrix}a-d, reveals some general behaviors. In the W-based nanobubbles, the number of forbidden excitations is lower than in their Mo-based counterparts. In the spectrum of the excitations from the flat state of the \ce{MoSe2} nanobubble at $F = 0.0175$~a.u./atom, the electronic structure, characterized by generally less dispersive and energetically more separated states compared to the other systems (compare \autoref{fig:BS-WFD-examples}e), leads to much less dark transitions than the other nanostructures of the same material (\autoref{fig:transition_matrix}c). Finally, a close inspection of \autoref{fig:transition_matrix}d reveals a regular pattern in the transition spectrum from the localized state in the \ce{MoS2} nanobubbles: regardless of the inflation force, all these excitations occurring from the mid-gap state to the lowest conduction states are forbidden.

Moving on to the spectra of transitions from the lowest dispersive valence band at $\Gamma$, we notice a more diversified scenario depending on the chemical composition of the nanobubbles and their inflating forces (\autoref{fig:transition_matrix}e-h). The W-based materials exhibit optically active low-energy transitions, targeting the second-lowest conduction state  at $\Gamma$, with oscillator strength decreasing with increasing inflating force (\autoref{fig:transition_matrix}e,f). \ce{MoSe2} nanobubbles subject to inflating forces from 0.01 to 0.015~a.u./atom are characterized by optically active transitions from the highest dispersive state, corresponding to the VBM (\autoref{fig:gapstates}b), to the CBm, see \autoref{fig:transition_matrix}g. Under larger deformations ($F = 0.0175$~a.u./atom and $F = 0.02$~a.u./atom), the spectrum becomes more similar to the one obtained for transitions from the flat state, due to the peculiar electronic structure reported in \autoref{fig:BS-WFD-examples}e. Finally, transitions from the highest dispersive valence state at $\Gamma$ to the lowest conduction states are generally very weak and even dark for the smallest deformations (e.g., $F = 0.01$~a.u./atom in \autoref{fig:transition_matrix}h). Active transitions appear higher in energy, around 2.5~eV and above.

\begin{figure}
    \centering
    \includegraphics[width=0.48\textwidth]{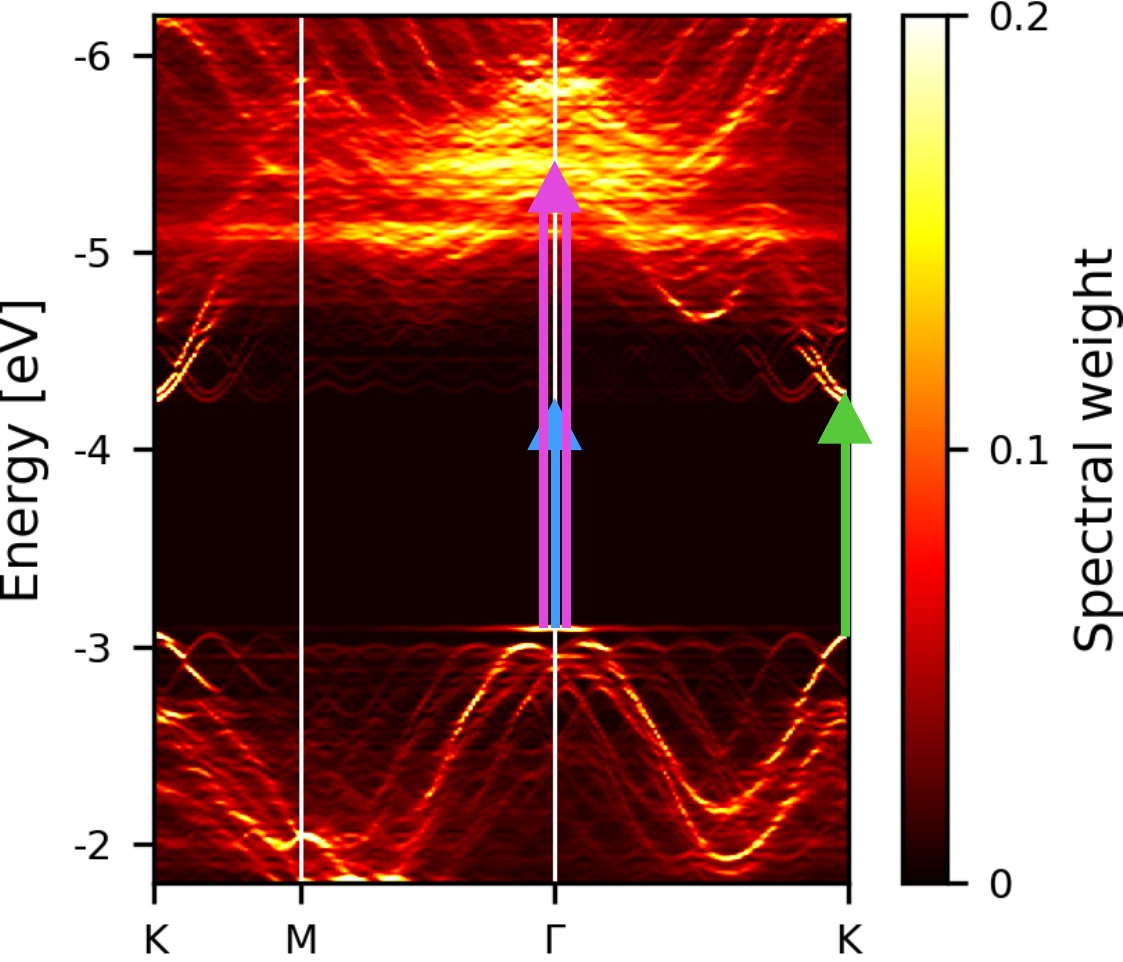}
    \caption{Unfolded band structure of the \ce{WS2} nanobubble at $F = 0.0175$~a.u./atom. The color scale marks the spectral weight associated with each electronic state. Dark and bright vertical transitions from the flat state at $\Gamma$ and the dispersive state at K are marked by the blue and green arrows, respectively. Allowed transitions from the flat mid-gap state to conduction bands between -3~eV and -2~eV are denoted by the magenta arrow. 
    }
    \label{fig:unfolded}
\end{figure}

The results discussed above suggest that the formation of a localized state at the top of the valence band of the considered TMDC nanobubbles does not enhance the optical yield compared to their flat counterparts. In contrast, the prediction of optically dark or very weak transitions from these states to the CBm discourages perspectives for single-photon emission. To better understand the origin of this unexpected behavior, we inspect the band structure of the \ce{WS2} nanobubble subject to $F = 0.0175$~a.u./atom unfolded into the hexagonal Brillouin zone of the pristine monolayer~\cite{krum-cocc21es}. 

As shown in \autoref{fig:unfolded}, the non-dispersive state is visible within the gap, but its spectral weight is focused around the $\Gamma$-point, fading toward the high-symmetry point K, where the valleys of the ideally flat material are located. In turn, the replicas of the K-valleys consistently decrease their spectral weight toward $\Gamma$ (blue arrow), giving rise to optically dark transitions from the flat state (\autoref{fig:transition_matrix}b). This distribution is caused by the curvature breaking the $\sigma_h$ symmetry, which rehybridizes the out-of-plane chalcogen $p_z$- with metal $d_{z^2}$-orbitals at $\Gamma$~\cite{krumland2024quantum}. The K point, instead, is dominated by in-plane orbitals, which participate significantly less in this rehybridization, thus causing the spectral weight of the flat state to fade in this region. The weak but non-forbidden transitions appearing at higher energies in \autoref{fig:transition_matrix}b emerge from the promotion of electrons from the flat mid-gap state to the unoccupied band manifold between -3~eV and -2~eV (magenta arrow in \autoref{fig:unfolded}), which are expected to participate in the formation of high-energy excitons in TMDCs, like C, D, or even beyond~\cite{gosw+21jpcl,li+21apl}. The most intense transitions giving rise to the lowest energy excitons in these monolayer materials appear at K (green arrow in \autoref{fig:unfolded}).
We emphasize that, due to breaking of the translation symmetry in the nanobubble compared to a pristine structure, $\*k$ is no longer a good quantum number and non-vertical optical transitions are, in principle, allowed. However, the symmetry breaking is not strong enough to significantly enable non-verticality of the optical transitions, which is a likely reason for them being dark. Thus, the unfolded band structure illustrates the physics in play, accompanying the oscillator strength results.

Overall, this analysis suggests that the $\mathbf{k}$-selection rules are likely the essential ingredient to explain the optical activity of vertical transitions from localized states in TMDC nanobubbles of the sub-10~nm size considered in this work. Our results reveal that these systems are not suited for single-photon emission at the apex, where such non-dispersive states are localized. To exclude artifacts due to the specific structure adopted to model these systems, especially regarding the fixed metal atoms at the margins of the supercell (\autoref{fig:structurePanels}a), we checked that modifying the extension of the region of clamped ions does not qualitatively change our findings. A reconciliation with experimental observations indicating TMDC nanobubbles as favorable candidates for quantum emission~\cite{darlington2020imaging,luo+18natn,blun+21prl,stel+24nl, cianci2023spatially} is possible only in terms of the sample size, which exceeds by a few orders of magnitude the one considered here. 

While spatial confinement in the considered sub-10-nm nanobubbles leads to quantum-dot-like flat states and the sizeable reduction of the bandgap size represents a favorable condition for exciton funneling, the resulting optical transitions remain weak due to the momentum mismatch between the $\Gamma$-localized valence state and the $K$-centered conduction band minimum. This indicates that the simple scaling down of nanobubbles to the sub-10-nm regime may not be sufficient for enhanced photon emission in pristine lattices, requiring instead the presence of defects or other extrinsic mechanisms modulating valley physics~\cite{krumland2024quantum}.
The (almost) mesoscopic scale of the experimentally realized TMDC nanobubbles~\cite{darlington2020imaging,blun+21prl} gives rise to quantum dots characterized by non-dispersive electronic states, which, as such, are insensitive to the valley localization of the CBm. In contrast, the model structures investigated in this work retain the periodicity of the monolayers, reflected in the characteristics of their electronic states and, consequently, of their optical transitions. 
The qualitative value of our predictions is confirmed by a comparison with experimental trends. The energy depth of our localized states (50--200~meV) aligns with the observed spectral separation of $\sim$100~meV  between localized and primary excitons in WSe$_2$ nanobubbles~\cite{darlington2020imaging}. Likewise, scanning tunneling spectroscopy measurements~\cite{shab+22nl} reveal localized electronic states and a gap reduction of $\sim$200~meV over a 15~nm scale.

Before concluding, it is worth stressing that the present calculations do not include a substrate, neither implicit nor explicit, in contrast to typical experimental setups \cite{cianci2023spatially, tedeschi2019controlled, blun+20prr, pucko2022excitons, di2022mechanical}. To support our conclusions despite this important missing ingredient, we can briefly discuss the expected impact of a substrate on the optical activity of low-dimensional systems like the considered TMDC nanobubbles. Screening is a ubiquitous effect that becomes dominant when the TMDCs are deposited on metals or strongly polarizable materials characterized by large dielectric functions. 
It reduces the quasiparticle gap while lowering exciton binding energies, often by a similar amount, which leads to at least a partial compensation well captured by bare DFT calculations due to error cancellation \cite{cho2018environmentally, raja2017coulomb, florian2018dielectric}. Effective models have proven very successful in describing the underlying physics at moderate computational costs~\cite{krumland2021layerpcm,tanda2024evidence,krumland2024electronic,krumland2024ab}. In contrast, chemical interactions with the substrate can cause profound structural and electronic-structure modifications in the adsorbate, requiring a detailed, atomistic description that is unaffordable within a quantum-mechanical formalism like DFT. However, this scenario can be considered unlikely for inflated TMDC nanobubbles, where the substantial deformation reduces chemical interactions.

\section{Summary and Conclusions}

In summary, we performed a comprehensive \textit{ab initio} investigation of defect-free TMDC nanobubbles of various compositions (MoS$_2$, WS$_2$, MoSe$_2$, WSe$_2$) subject to a set of inflating forces, systematically exploring the interplay between their structural and electronic properties, and the consequences for their optical activity. The applied structural modifications lead to geometries with varying heights and curvatures. The induced strain distribution is highly non-uniform, with maximum tensile strain concentrated at the apex. Material-specific differences in rigidity play a crucial role, with MoS$_2$ exhibiting the highest stiffness and WSe$_2$ the largest flexibility, influencing the maximum achievable strain and the resulting bubble morphology.
The electronic properties of the nanobubbles are significantly modified by strain. All band gaps exhibit a pseudoparabolic decrease with increasing inflation, with trends at weak forces mirroring those of flat monolayers. 

A non-dispersive (flat) electronic state in the valence region appears in all considered TMDC nanobubbles. For MoS$_2$ and WS$_2$, these states are driven into the fundamental gap at higher forces, while for MoSe$_2$ and WSe$_2$, they remain within the valence band, albeit close to the VBM. Quantification of their wave-function distribution revealed distinct, material-dependent localization behaviors. For MoS$_2$ and WS$_2$, the VBM becomes highly localized at the apex with increasing force. For MoSe$_2$ and WSe$_2$, while the VBM shifts its localization away from the apex towards the flat supercell edges, specific underlying localized valence states exhibit a substantial and increasing localization at the apex. The conduction band minimum, instead, generally remains insensitive to strain, maintaining a largely delocalized character. The orbital character of these frontier states also shows sensitivity to force and composition, with metal $d$- and upper chalcogen p-orbital contributions varying with chemical composition according to the deformation-induced broken mirror symmetry. 
Crucially, transitions from the apex-localized states (including those in the mid-gap for MoS$_2$ and WS$_2$) to the CBm are predominantly dark. This lack of optical activity is attributed to the persistent localization of these valence states at $\Gamma$, in contrast to the K-valley localization of the CBm, as revealed by the analysis of the unfolded band structure. This condition weakens optically active vertical transitions from the non-dispersive state in the mid-gap, hindering the potential of the considered TMDC nanobubbles as single-photon emitters. 

In conclusion, our results suggest that while mechanical inflation successfully creates localized electronic states in defect-free TMDC nanobubbles, the latter generally lack the intrinsic optical activity required for efficient single-photon emission. Our outcomes imply that this phenomenon, observed in significantly larger nanobubbles~\cite{darlington2020imaging,luo+18natn,blun+21prl,stel+24nl}, and predicted for nanobubbles the size of ours \cite{krumland2024quantum}, also requires a deep, discrete exciton trap, or sharp deformations causing CBm localization, respectively. This insight, while delivering an apparent negative result, is paramount for comprehensively understanding strain-tailored TMDC nanostructures. 
Upcoming theoretical and computational efforts in this field should thus extend the focus to more realistic structures, incorporating defects~\cite{krumland2024quantum} and systematically including excitonic effects, bridging the current gap between pure theoretical models and experimental observations of quantum emitters. This will be key to exploiting their full potential in novel quantum applications.

\section*{Acknowledgements}
We thank Surender Kumar for fruitful discussions.
This work was funded by the QuanterERA II European Union’s Horizon 2020 research and innovation program under the EQUAISE project, Grant Agreement no. 101017733, by the German Federal Ministry of Education and Research (Professorinnenprogramm III), and by the State of Lower Saxony (Professorinnen f\"ur Niedersachsen, SMART, and DyNano). The authors gratefully acknowledge the computing time made available to them on the high-performance computer Emmy in G\"ottingen, supported by the German Federal Ministry of Education and Research and the state governments participating in the National High-Performance Computing (NHR) funding program.

\section*{Data Availability}
The datasets generated and analyzed during the current study are available in Zenodo, DOI: 10.5281/zenodo.15830853.


\begin{thebibliography}{75}%
\makeatletter
\providecommand \@ifxundefined [1]{%
 \@ifx{#1\undefined}
}%
\providecommand \@ifnum [1]{%
 \ifnum #1\expandafter \@firstoftwo
 \else \expandafter \@secondoftwo
 \fi
}%
\providecommand \@ifx [1]{%
 \ifx #1\expandafter \@firstoftwo
 \else \expandafter \@secondoftwo
 \fi
}%
\providecommand \natexlab [1]{#1}%
\providecommand \enquote  [1]{``#1''}%
\providecommand \bibnamefont  [1]{#1}%
\providecommand \bibfnamefont [1]{#1}%
\providecommand \citenamefont [1]{#1}%
\providecommand \href@noop [0]{\@secondoftwo}%
\providecommand \href [0]{\begingroup \@sanitize@url \@href}%
\providecommand \@href[1]{\@@startlink{#1}\@@href}%
\providecommand \@@href[1]{\endgroup#1\@@endlink}%
\providecommand \@sanitize@url [0]{\catcode `\\12\catcode `\$12\catcode
  `\&12\catcode `\#12\catcode `\^12\catcode `\_12\catcode `\%12\relax}%
\providecommand \@@startlink[1]{}%
\providecommand \@@endlink[0]{}%
\providecommand \url  [0]{\begingroup\@sanitize@url \@url }%
\providecommand \@url [1]{\endgroup\@href {#1}{\urlprefix }}%
\providecommand \urlprefix  [0]{URL }%
\providecommand \Eprint [0]{\href }%
\providecommand \doibase [0]{https://doi.org/}%
\providecommand \selectlanguage [0]{\@gobble}%
\providecommand \bibinfo  [0]{\@secondoftwo}%
\providecommand \bibfield  [0]{\@secondoftwo}%
\providecommand \translation [1]{[#1]}%
\providecommand \BibitemOpen [0]{}%
\providecommand \bibitemStop [0]{}%
\providecommand \bibitemNoStop [0]{.\EOS\space}%
\providecommand \EOS [0]{\spacefactor3000\relax}%
\providecommand \BibitemShut  [1]{\csname bibitem#1\endcsname}%
\let\auto@bib@innerbib\@empty
\bibitem [{\citenamefont {Levy}(2022)}]{levy22aplm}%
  \BibitemOpen
  \bibfield  {author} {\bibinfo {author} {\bibfnamefont {J.}~\bibnamefont
  {Levy}},\ }\bibfield  {title} {\bibinfo {title} {Correlated nanoelectronics
  and the second quantum revolution},\ }\href@noop {} {\bibfield  {journal}
  {\bibinfo  {journal} {APL~Mater.}\ }\textbf {\bibinfo {volume} {10}},\
  \bibinfo {pages} {110901} (\bibinfo {year} {2022})}\BibitemShut {NoStop}%
\bibitem [{\citenamefont {Coleman}\ \emph {et~al.}(2011)\citenamefont
  {Coleman}, \citenamefont {Lotya}, \citenamefont {O’Neill}, \citenamefont
  {Bergin}, \citenamefont {King}, \citenamefont {Khan}, \citenamefont {Young},
  \citenamefont {Gaucher}, \citenamefont {De}, \citenamefont {Smith} \emph
  {et~al.}}]{cole+11sci}%
  \BibitemOpen
  \bibfield  {author} {\bibinfo {author} {\bibfnamefont {J.~N.}\ \bibnamefont
  {Coleman}}, \bibinfo {author} {\bibfnamefont {M.}~\bibnamefont {Lotya}},
  \bibinfo {author} {\bibfnamefont {A.}~\bibnamefont {O’Neill}}, \bibinfo
  {author} {\bibfnamefont {S.~D.}\ \bibnamefont {Bergin}}, \bibinfo {author}
  {\bibfnamefont {P.~J.}\ \bibnamefont {King}}, \bibinfo {author}
  {\bibfnamefont {U.}~\bibnamefont {Khan}}, \bibinfo {author} {\bibfnamefont
  {K.}~\bibnamefont {Young}}, \bibinfo {author} {\bibfnamefont
  {A.}~\bibnamefont {Gaucher}}, \bibinfo {author} {\bibfnamefont
  {S.}~\bibnamefont {De}}, \bibinfo {author} {\bibfnamefont {R.~J.}\
  \bibnamefont {Smith}}, \emph {et~al.},\ }\bibfield  {title} {\bibinfo {title}
  {Two-dimensional nanosheets produced by liquid exfoliation of layered
  materials},\ }\href@noop {} {\bibfield  {journal} {\bibinfo  {journal}
  {Science}\ }\textbf {\bibinfo {volume} {331}},\ \bibinfo {pages} {568}
  (\bibinfo {year} {2011})}\BibitemShut {NoStop}%
\bibitem [{\citenamefont {Liu}\ and\ \citenamefont
  {Hersam}(2019)}]{liu-hers19nrm}%
  \BibitemOpen
  \bibfield  {author} {\bibinfo {author} {\bibfnamefont {X.}~\bibnamefont
  {Liu}}\ and\ \bibinfo {author} {\bibfnamefont {M.~C.}\ \bibnamefont
  {Hersam}},\ }\bibfield  {title} {\bibinfo {title} {{2D} materials for quantum
  information science},\ }\href@noop {} {\bibfield  {journal} {\bibinfo
  {journal} {Nat.~Rev.~Mater.}\ }\textbf {\bibinfo {volume} {4}},\ \bibinfo
  {pages} {669} (\bibinfo {year} {2019})}\BibitemShut {NoStop}%
\bibitem [{\citenamefont {Fiori}\ \emph {et~al.}(2014)\citenamefont {Fiori},
  \citenamefont {Bonaccorso}, \citenamefont {Iannaccone}, \citenamefont
  {Palacios}, \citenamefont {Neumaier}, \citenamefont {Seabaugh}, \citenamefont
  {Banerjee},\ and\ \citenamefont {Colombo}}]{fior+14natn}%
  \BibitemOpen
  \bibfield  {author} {\bibinfo {author} {\bibfnamefont {G.}~\bibnamefont
  {Fiori}}, \bibinfo {author} {\bibfnamefont {F.}~\bibnamefont {Bonaccorso}},
  \bibinfo {author} {\bibfnamefont {G.}~\bibnamefont {Iannaccone}}, \bibinfo
  {author} {\bibfnamefont {T.}~\bibnamefont {Palacios}}, \bibinfo {author}
  {\bibfnamefont {D.}~\bibnamefont {Neumaier}}, \bibinfo {author}
  {\bibfnamefont {A.}~\bibnamefont {Seabaugh}}, \bibinfo {author}
  {\bibfnamefont {S.~K.}\ \bibnamefont {Banerjee}},\ and\ \bibinfo {author}
  {\bibfnamefont {L.}~\bibnamefont {Colombo}},\ }\bibfield  {title} {\bibinfo
  {title} {Electronics based on two-dimensional materials},\ }\href@noop {}
  {\bibfield  {journal} {\bibinfo  {journal} {Nature~Nanotechnol.}\ }\textbf
  {\bibinfo {volume} {9}},\ \bibinfo {pages} {768} (\bibinfo {year}
  {2014})}\BibitemShut {NoStop}%
\bibitem [{\citenamefont {Wang}\ \emph {et~al.}(2018)\citenamefont {Wang},
  \citenamefont {Chernikov}, \citenamefont {Glazov}, \citenamefont {Heinz},
  \citenamefont {Marie}, \citenamefont {Amand},\ and\ \citenamefont
  {Urbaszek}}]{wang+18rmp}%
  \BibitemOpen
  \bibfield  {author} {\bibinfo {author} {\bibfnamefont {G.}~\bibnamefont
  {Wang}}, \bibinfo {author} {\bibfnamefont {A.}~\bibnamefont {Chernikov}},
  \bibinfo {author} {\bibfnamefont {M.~M.}\ \bibnamefont {Glazov}}, \bibinfo
  {author} {\bibfnamefont {T.~F.}\ \bibnamefont {Heinz}}, \bibinfo {author}
  {\bibfnamefont {X.}~\bibnamefont {Marie}}, \bibinfo {author} {\bibfnamefont
  {T.}~\bibnamefont {Amand}},\ and\ \bibinfo {author} {\bibfnamefont
  {B.}~\bibnamefont {Urbaszek}},\ }\bibfield  {title} {\bibinfo {title}
  {Colloquium: Excitons in atomically thin transition metal dichalcogenides},\
  }\href@noop {} {\bibfield  {journal} {\bibinfo  {journal} {Rev.~Mod.~Phys.}\
  }\textbf {\bibinfo {volume} {90}},\ \bibinfo {pages} {021001} (\bibinfo
  {year} {2018})}\BibitemShut {NoStop}%
\bibitem [{\citenamefont {Avsar}\ \emph {et~al.}(2020)\citenamefont {Avsar},
  \citenamefont {Ochoa}, \citenamefont {Guinea}, \citenamefont {{\"O}zyilmaz},
  \citenamefont {Van~Wees},\ and\ \citenamefont {Vera-Marun}}]{avsa+20rmp}%
  \BibitemOpen
  \bibfield  {author} {\bibinfo {author} {\bibfnamefont {A.}~\bibnamefont
  {Avsar}}, \bibinfo {author} {\bibfnamefont {H.}~\bibnamefont {Ochoa}},
  \bibinfo {author} {\bibfnamefont {F.}~\bibnamefont {Guinea}}, \bibinfo
  {author} {\bibfnamefont {B.}~\bibnamefont {{\"O}zyilmaz}}, \bibinfo {author}
  {\bibfnamefont {B.}~\bibnamefont {Van~Wees}},\ and\ \bibinfo {author}
  {\bibfnamefont {I.~J.}\ \bibnamefont {Vera-Marun}},\ }\bibfield  {title}
  {\bibinfo {title} {Colloquium: Spintronics in graphene and other
  two-dimensional materials},\ }\href@noop {} {\bibfield  {journal} {\bibinfo
  {journal} {Rev.~Mod.~Phys.}\ }\textbf {\bibinfo {volume} {92}},\ \bibinfo
  {pages} {021003} (\bibinfo {year} {2020})}\BibitemShut {NoStop}%
\bibitem [{\citenamefont {Duerloo}\ \emph {et~al.}(2014)\citenamefont
  {Duerloo}, \citenamefont {Li},\ and\ \citenamefont {Reed}}]{duer+14natcom}%
  \BibitemOpen
  \bibfield  {author} {\bibinfo {author} {\bibfnamefont {K.-A.~N.}\
  \bibnamefont {Duerloo}}, \bibinfo {author} {\bibfnamefont {Y.}~\bibnamefont
  {Li}},\ and\ \bibinfo {author} {\bibfnamefont {E.~J.}\ \bibnamefont {Reed}},\
  }\bibfield  {title} {\bibinfo {title} {Structural phase transitions in
  two-dimensional {Mo}-and {W}-dichalcogenide monolayers},\ }\href@noop {}
  {\bibfield  {journal} {\bibinfo  {journal} {Nat.~Commun.}\ }\textbf {\bibinfo
  {volume} {5}},\ \bibinfo {pages} {4214} (\bibinfo {year} {2014})}\BibitemShut
  {NoStop}%
\bibitem [{\citenamefont {Song}\ \emph {et~al.}(2016)\citenamefont {Song},
  \citenamefont {Keum}, \citenamefont {Cho}, \citenamefont {Perello},
  \citenamefont {Kim},\ and\ \citenamefont {Lee}}]{song+16nl}%
  \BibitemOpen
  \bibfield  {author} {\bibinfo {author} {\bibfnamefont {S.}~\bibnamefont
  {Song}}, \bibinfo {author} {\bibfnamefont {D.~H.}\ \bibnamefont {Keum}},
  \bibinfo {author} {\bibfnamefont {S.}~\bibnamefont {Cho}}, \bibinfo {author}
  {\bibfnamefont {D.}~\bibnamefont {Perello}}, \bibinfo {author} {\bibfnamefont
  {Y.}~\bibnamefont {Kim}},\ and\ \bibinfo {author} {\bibfnamefont {Y.~H.}\
  \bibnamefont {Lee}},\ }\bibfield  {title} {\bibinfo {title} {Room temperature
  semiconductor--metal transition of \ce{MoTe2} thin films engineered by
  strain},\ }\href@noop {} {\bibfield  {journal} {\bibinfo  {journal}
  {Nano~Lett.}\ }\textbf {\bibinfo {volume} {16}},\ \bibinfo {pages} {188}
  (\bibinfo {year} {2016})}\BibitemShut {NoStop}%
\bibitem [{\citenamefont {Maiti}\ \emph {et~al.}(2020)\citenamefont {Maiti},
  \citenamefont {Patil}, \citenamefont {Saadi}, \citenamefont {Xie},
  \citenamefont {Azadani}, \citenamefont {Uluutku}, \citenamefont {Amin},
  \citenamefont {Briggs}, \citenamefont {Miscuglio}, \citenamefont
  {Van~Thourhout} \emph {et~al.}}]{mait+20natph}%
  \BibitemOpen
  \bibfield  {author} {\bibinfo {author} {\bibfnamefont {R.}~\bibnamefont
  {Maiti}}, \bibinfo {author} {\bibfnamefont {C.}~\bibnamefont {Patil}},
  \bibinfo {author} {\bibfnamefont {M.}~\bibnamefont {Saadi}}, \bibinfo
  {author} {\bibfnamefont {T.}~\bibnamefont {Xie}}, \bibinfo {author}
  {\bibfnamefont {J.}~\bibnamefont {Azadani}}, \bibinfo {author} {\bibfnamefont
  {B.}~\bibnamefont {Uluutku}}, \bibinfo {author} {\bibfnamefont
  {R.}~\bibnamefont {Amin}}, \bibinfo {author} {\bibfnamefont {A.}~\bibnamefont
  {Briggs}}, \bibinfo {author} {\bibfnamefont {M.}~\bibnamefont {Miscuglio}},
  \bibinfo {author} {\bibfnamefont {D.}~\bibnamefont {Van~Thourhout}}, \emph
  {et~al.},\ }\bibfield  {title} {\bibinfo {title} {Strain-engineered
  high-responsivity \ce{MoTe2} photodetector for silicon photonic integrated
  circuits},\ }\href@noop {} {\bibfield  {journal} {\bibinfo  {journal}
  {Nature~Photon.}\ }\textbf {\bibinfo {volume} {14}},\ \bibinfo {pages} {578}
  (\bibinfo {year} {2020})}\BibitemShut {NoStop}%
\bibitem [{\citenamefont {Darlington}\ \emph {et~al.}(2020)\citenamefont
  {Darlington}, \citenamefont {Carmesin}, \citenamefont {Florian},
  \citenamefont {Yanev}, \citenamefont {Ajayi}, \citenamefont {Ardelean},
  \citenamefont {Rhodes}, \citenamefont {Ghiotto}, \citenamefont {Krayev},
  \citenamefont {Watanabe} \emph {et~al.}}]{darlington2020imaging}%
  \BibitemOpen
  \bibfield  {author} {\bibinfo {author} {\bibfnamefont {T.~P.}\ \bibnamefont
  {Darlington}}, \bibinfo {author} {\bibfnamefont {C.}~\bibnamefont
  {Carmesin}}, \bibinfo {author} {\bibfnamefont {M.}~\bibnamefont {Florian}},
  \bibinfo {author} {\bibfnamefont {E.}~\bibnamefont {Yanev}}, \bibinfo
  {author} {\bibfnamefont {O.}~\bibnamefont {Ajayi}}, \bibinfo {author}
  {\bibfnamefont {J.}~\bibnamefont {Ardelean}}, \bibinfo {author}
  {\bibfnamefont {D.~A.}\ \bibnamefont {Rhodes}}, \bibinfo {author}
  {\bibfnamefont {A.}~\bibnamefont {Ghiotto}}, \bibinfo {author} {\bibfnamefont
  {A.}~\bibnamefont {Krayev}}, \bibinfo {author} {\bibfnamefont
  {K.}~\bibnamefont {Watanabe}}, \emph {et~al.},\ }\bibfield  {title} {\bibinfo
  {title} {Imaging strain-localized excitons in nanoscale bubbles of monolayer
  \ce{WSe2} at room temperature},\ }\href@noop {} {\bibfield  {journal}
  {\bibinfo  {journal} {Nature~Nanotechnol.}\ }\textbf {\bibinfo {volume}
  {15}},\ \bibinfo {pages} {854} (\bibinfo {year} {2020})}\BibitemShut
  {NoStop}%
\bibitem [{\citenamefont {Blundo}\ \emph {et~al.}(2020)\citenamefont {Blundo},
  \citenamefont {Felici}, \citenamefont {Yildirim}, \citenamefont {Pettinari},
  \citenamefont {Tedeschi}, \citenamefont {Miriametro}, \citenamefont {Liu},
  \citenamefont {Ma}, \citenamefont {Lu},\ and\ \citenamefont
  {Polimeni}}]{blun+20prr}%
  \BibitemOpen
  \bibfield  {author} {\bibinfo {author} {\bibfnamefont {E.}~\bibnamefont
  {Blundo}}, \bibinfo {author} {\bibfnamefont {M.}~\bibnamefont {Felici}},
  \bibinfo {author} {\bibfnamefont {T.}~\bibnamefont {Yildirim}}, \bibinfo
  {author} {\bibfnamefont {G.}~\bibnamefont {Pettinari}}, \bibinfo {author}
  {\bibfnamefont {D.}~\bibnamefont {Tedeschi}}, \bibinfo {author}
  {\bibfnamefont {A.}~\bibnamefont {Miriametro}}, \bibinfo {author}
  {\bibfnamefont {B.}~\bibnamefont {Liu}}, \bibinfo {author} {\bibfnamefont
  {W.}~\bibnamefont {Ma}}, \bibinfo {author} {\bibfnamefont {Y.}~\bibnamefont
  {Lu}},\ and\ \bibinfo {author} {\bibfnamefont {A.}~\bibnamefont {Polimeni}},\
  }\bibfield  {title} {\bibinfo {title} {Evidence of the direct-to-indirect
  band gap transition in strained two-dimensional \ce{WS2}, \ce{MoS2}, and
  \ce{WSe2}},\ }\href@noop {} {\bibfield  {journal} {\bibinfo  {journal}
  {Phys.~Rev.~Res.}\ }\textbf {\bibinfo {volume} {2}},\ \bibinfo {pages}
  {012024} (\bibinfo {year} {2020})}\BibitemShut {NoStop}%
\bibitem [{\citenamefont {Blundo}\ \emph {et~al.}(2021)\citenamefont {Blundo},
  \citenamefont {Yildirim}, \citenamefont {Pettinari},\ and\ \citenamefont
  {Polimeni}}]{blun+21prl}%
  \BibitemOpen
  \bibfield  {author} {\bibinfo {author} {\bibfnamefont {E.}~\bibnamefont
  {Blundo}}, \bibinfo {author} {\bibfnamefont {T.}~\bibnamefont {Yildirim}},
  \bibinfo {author} {\bibfnamefont {G.}~\bibnamefont {Pettinari}},\ and\
  \bibinfo {author} {\bibfnamefont {A.}~\bibnamefont {Polimeni}},\ }\bibfield
  {title} {\bibinfo {title} {Experimental adhesion energy in van der {Waals}
  crystals and heterostructures from atomically thin bubbles},\ }\href
  {https://doi.org/10.1103/PhysRevLett.127.046101} {\bibfield  {journal}
  {\bibinfo  {journal} {Phys.~Rev.~Lett.}\ }\textbf {\bibinfo {volume} {127}},\
  \bibinfo {pages} {046101} (\bibinfo {year} {2021})}\BibitemShut {NoStop}%
\bibitem [{\citenamefont {Stellino}\ \emph {et~al.}(2024)\citenamefont
  {Stellino}, \citenamefont {D’Alo}, \citenamefont {Blundo}, \citenamefont
  {Postorino},\ and\ \citenamefont {Polimeni}}]{stel+24nl}%
  \BibitemOpen
  \bibfield  {author} {\bibinfo {author} {\bibfnamefont {E.}~\bibnamefont
  {Stellino}}, \bibinfo {author} {\bibfnamefont {B.}~\bibnamefont {D’Alo}},
  \bibinfo {author} {\bibfnamefont {E.}~\bibnamefont {Blundo}}, \bibinfo
  {author} {\bibfnamefont {P.}~\bibnamefont {Postorino}},\ and\ \bibinfo
  {author} {\bibfnamefont {A.}~\bibnamefont {Polimeni}},\ }\bibfield  {title}
  {\bibinfo {title} {Fine-tuning of the excitonic response in monolayer
  \ce{WS2} domes via coupled pressure and strain variation},\ }\href@noop {}
  {\bibfield  {journal} {\bibinfo  {journal} {Nano~Lett.}\ }\textbf {\bibinfo
  {volume} {24}},\ \bibinfo {pages} {3945} (\bibinfo {year}
  {2024})}\BibitemShut {NoStop}%
\bibitem [{\citenamefont {Parto}\ \emph {et~al.}(2021)\citenamefont {Parto},
  \citenamefont {Azzam}, \citenamefont {Banerjee},\ and\ \citenamefont
  {Moody}}]{part+21natcom}%
  \BibitemOpen
  \bibfield  {author} {\bibinfo {author} {\bibfnamefont {K.}~\bibnamefont
  {Parto}}, \bibinfo {author} {\bibfnamefont {S.~I.}\ \bibnamefont {Azzam}},
  \bibinfo {author} {\bibfnamefont {K.}~\bibnamefont {Banerjee}},\ and\
  \bibinfo {author} {\bibfnamefont {G.}~\bibnamefont {Moody}},\ }\bibfield
  {title} {\bibinfo {title} {Defect and strain engineering of monolayer
  \ce{WSe2} enables site-controlled single-photon emission up to 150 {K}},\
  }\href@noop {} {\bibfield  {journal} {\bibinfo  {journal} {Nat.~Commun.}\
  }\textbf {\bibinfo {volume} {12}},\ \bibinfo {pages} {3585} (\bibinfo {year}
  {2021})}\BibitemShut {NoStop}%
\bibitem [{\citenamefont {Kim}\ \emph {et~al.}(2022)\citenamefont {Kim},
  \citenamefont {Kim}, \citenamefont {Kumar}, \citenamefont {Rahaman},
  \citenamefont {Stevens}, \citenamefont {Jeon}, \citenamefont {Jo},
  \citenamefont {Kim}, \citenamefont {Trainor}, \citenamefont {Zhu} \emph
  {et~al.}}]{kim+22nano}%
  \BibitemOpen
  \bibfield  {author} {\bibinfo {author} {\bibfnamefont {G.}~\bibnamefont
  {Kim}}, \bibinfo {author} {\bibfnamefont {H.~M.}\ \bibnamefont {Kim}},
  \bibinfo {author} {\bibfnamefont {P.}~\bibnamefont {Kumar}}, \bibinfo
  {author} {\bibfnamefont {M.}~\bibnamefont {Rahaman}}, \bibinfo {author}
  {\bibfnamefont {C.~E.}\ \bibnamefont {Stevens}}, \bibinfo {author}
  {\bibfnamefont {J.}~\bibnamefont {Jeon}}, \bibinfo {author} {\bibfnamefont
  {K.}~\bibnamefont {Jo}}, \bibinfo {author} {\bibfnamefont {K.-H.}\
  \bibnamefont {Kim}}, \bibinfo {author} {\bibfnamefont {N.}~\bibnamefont
  {Trainor}}, \bibinfo {author} {\bibfnamefont {H.}~\bibnamefont {Zhu}}, \emph
  {et~al.},\ }\bibfield  {title} {\bibinfo {title} {High-density, localized
  quantum emitters in strained {2D} semiconductors},\ }\href@noop {} {\bibfield
   {journal} {\bibinfo  {journal} {ACS~Nano}\ }\textbf {\bibinfo {volume}
  {16}},\ \bibinfo {pages} {9651} (\bibinfo {year} {2022})}\BibitemShut
  {NoStop}%
\bibitem [{\citenamefont {Li}\ \emph {et~al.}(2022)\citenamefont {Li},
  \citenamefont {Chui}, \citenamefont {Shen}, \citenamefont {Huang},
  \citenamefont {Wen}, \citenamefont {Yam}, \citenamefont {Shao}, \citenamefont
  {Xu},\ and\ \citenamefont {Wang}}]{li+22nano}%
  \BibitemOpen
  \bibfield  {author} {\bibinfo {author} {\bibfnamefont {S.}~\bibnamefont
  {Li}}, \bibinfo {author} {\bibfnamefont {K.~K.}\ \bibnamefont {Chui}},
  \bibinfo {author} {\bibfnamefont {F.}~\bibnamefont {Shen}}, \bibinfo {author}
  {\bibfnamefont {H.}~\bibnamefont {Huang}}, \bibinfo {author} {\bibfnamefont
  {S.}~\bibnamefont {Wen}}, \bibinfo {author} {\bibfnamefont {C.}~\bibnamefont
  {Yam}}, \bibinfo {author} {\bibfnamefont {L.}~\bibnamefont {Shao}}, \bibinfo
  {author} {\bibfnamefont {J.}~\bibnamefont {Xu}},\ and\ \bibinfo {author}
  {\bibfnamefont {J.}~\bibnamefont {Wang}},\ }\bibfield  {title} {\bibinfo
  {title} {Generation and detection of strain-localized excitons in \ce{WS2}
  monolayer by plasmonic metal nanocrystals},\ }\href@noop {} {\bibfield
  {journal} {\bibinfo  {journal} {ACS~Nano}\ }\textbf {\bibinfo {volume}
  {16}},\ \bibinfo {pages} {10647} (\bibinfo {year} {2022})}\BibitemShut
  {NoStop}%
\bibitem [{\citenamefont {Shabani}\ \emph {et~al.}(2022)\citenamefont
  {Shabani}, \citenamefont {Darlington}, \citenamefont {Gordon}, \citenamefont
  {Wu}, \citenamefont {Yanev}, \citenamefont {Hone}, \citenamefont {Zhu},
  \citenamefont {Dreyer}, \citenamefont {Schuck},\ and\ \citenamefont
  {Pasupathy}}]{shab+22nl}%
  \BibitemOpen
  \bibfield  {author} {\bibinfo {author} {\bibfnamefont {S.}~\bibnamefont
  {Shabani}}, \bibinfo {author} {\bibfnamefont {T.~P.}\ \bibnamefont
  {Darlington}}, \bibinfo {author} {\bibfnamefont {C.}~\bibnamefont {Gordon}},
  \bibinfo {author} {\bibfnamefont {W.}~\bibnamefont {Wu}}, \bibinfo {author}
  {\bibfnamefont {E.}~\bibnamefont {Yanev}}, \bibinfo {author} {\bibfnamefont
  {J.}~\bibnamefont {Hone}}, \bibinfo {author} {\bibfnamefont {X.}~\bibnamefont
  {Zhu}}, \bibinfo {author} {\bibfnamefont {C.~E.}\ \bibnamefont {Dreyer}},
  \bibinfo {author} {\bibfnamefont {P.~J.}\ \bibnamefont {Schuck}},\ and\
  \bibinfo {author} {\bibfnamefont {A.~N.}\ \bibnamefont {Pasupathy}},\
  }\bibfield  {title} {\bibinfo {title} {Ultralocalized optoelectronic
  properties of nanobubbles in {2D} semiconductors},\ }\href@noop {} {\bibfield
   {journal} {\bibinfo  {journal} {Nano~Lett.}\ }\textbf {\bibinfo {volume}
  {22}},\ \bibinfo {pages} {7401} (\bibinfo {year} {2022})}\BibitemShut
  {NoStop}%
\bibitem [{\citenamefont {Azzam}\ \emph {et~al.}(2021)\citenamefont {Azzam},
  \citenamefont {Parto},\ and\ \citenamefont {Moody}}]{azza+21apl}%
  \BibitemOpen
  \bibfield  {author} {\bibinfo {author} {\bibfnamefont {S.~I.}\ \bibnamefont
  {Azzam}}, \bibinfo {author} {\bibfnamefont {K.}~\bibnamefont {Parto}},\ and\
  \bibinfo {author} {\bibfnamefont {G.}~\bibnamefont {Moody}},\ }\bibfield
  {title} {\bibinfo {title} {Prospects and challenges of quantum emitters in
  {2D} materials},\ }\href@noop {} {\bibfield  {journal} {\bibinfo  {journal}
  {Appl.~Phys.~Lett.}\ }\textbf {\bibinfo {volume} {118}},\ \bibinfo {pages}
  {240502} (\bibinfo {year} {2021})}\BibitemShut {NoStop}%
\bibitem [{\citenamefont {Yu}\ \emph {et~al.}(2025)\citenamefont {Yu},
  \citenamefont {Ge}, \citenamefont {Luo}, \citenamefont {Seo}, \citenamefont
  {Kim}, \citenamefont {Eng}, \citenamefont {Lu}, \citenamefont {Wei},
  \citenamefont {Lee}, \citenamefont {Gao} \emph {et~al.}}]{yu+25nl}%
  \BibitemOpen
  \bibfield  {author} {\bibinfo {author} {\bibfnamefont {Y.}~\bibnamefont
  {Yu}}, \bibinfo {author} {\bibfnamefont {J.}~\bibnamefont {Ge}}, \bibinfo
  {author} {\bibfnamefont {M.}~\bibnamefont {Luo}}, \bibinfo {author}
  {\bibfnamefont {I.~C.}\ \bibnamefont {Seo}}, \bibinfo {author} {\bibfnamefont
  {Y.}~\bibnamefont {Kim}}, \bibinfo {author} {\bibfnamefont {J.~J.}\
  \bibnamefont {Eng}}, \bibinfo {author} {\bibfnamefont {K.}~\bibnamefont
  {Lu}}, \bibinfo {author} {\bibfnamefont {T.-R.}\ \bibnamefont {Wei}},
  \bibinfo {author} {\bibfnamefont {S.~W.}\ \bibnamefont {Lee}}, \bibinfo
  {author} {\bibfnamefont {W.}~\bibnamefont {Gao}}, \emph {et~al.},\ }\bibfield
   {title} {\bibinfo {title} {Dynamic tuning of single-photon emission in
  monolayer wse$_2$ via localized strain engineering},\ }\href@noop {}
  {\bibfield  {journal} {\bibinfo  {journal} {Nano~Lett.}\ }\textbf {\bibinfo
  {volume} {25}},\ \bibinfo {pages} {3438} (\bibinfo {year}
  {2025})}\BibitemShut {NoStop}%
\bibitem [{\citenamefont {Blundo}\ \emph {et~al.}(2025)\citenamefont {Blundo},
  \citenamefont {Tuzi}, \citenamefont {Cuccu}, \citenamefont {Re~Fiorentin},
  \citenamefont {Pettinari}, \citenamefont {Patra}, \citenamefont {Cianci},
  \citenamefont {Kudrynskyi}, \citenamefont {Felici}, \citenamefont {Taniguchi}
  \emph {et~al.}}]{blun+25nl}%
  \BibitemOpen
  \bibfield  {author} {\bibinfo {author} {\bibfnamefont {E.}~\bibnamefont
  {Blundo}}, \bibinfo {author} {\bibfnamefont {F.}~\bibnamefont {Tuzi}},
  \bibinfo {author} {\bibfnamefont {M.}~\bibnamefont {Cuccu}}, \bibinfo
  {author} {\bibfnamefont {M.}~\bibnamefont {Re~Fiorentin}}, \bibinfo {author}
  {\bibfnamefont {G.}~\bibnamefont {Pettinari}}, \bibinfo {author}
  {\bibfnamefont {A.}~\bibnamefont {Patra}}, \bibinfo {author} {\bibfnamefont
  {S.}~\bibnamefont {Cianci}}, \bibinfo {author} {\bibfnamefont {Z.~R.}\
  \bibnamefont {Kudrynskyi}}, \bibinfo {author} {\bibfnamefont
  {M.}~\bibnamefont {Felici}}, \bibinfo {author} {\bibfnamefont
  {T.}~\bibnamefont {Taniguchi}}, \emph {et~al.},\ }\bibfield  {title}
  {\bibinfo {title} {Giant light emission enhancement in strain-engineered
  {InSe}/{MS}$_2$ ({M}= {Mo} or {W}) van der {Waals} heterostructures},\
  }\href@noop {} {\bibfield  {journal} {\bibinfo  {journal} {Nano~Lett.}\
  }\textbf {\bibinfo {volume} {25}},\ \bibinfo {pages} {3375} (\bibinfo {year}
  {2025})}\BibitemShut {NoStop}%
\bibitem [{\citenamefont {Chen}\ \emph {et~al.}(2022)\citenamefont {Chen},
  \citenamefont {Song}, \citenamefont {Wang}, \citenamefont {Zhang},
  \citenamefont {Hu}, \citenamefont {Tian}, \citenamefont {Su}, \citenamefont
  {Chu}, \citenamefont {Zhang},\ and\ \citenamefont {Di}}]{chen+22ami}%
  \BibitemOpen
  \bibfield  {author} {\bibinfo {author} {\bibfnamefont {Y.}~\bibnamefont
  {Chen}}, \bibinfo {author} {\bibfnamefont {P.}~\bibnamefont {Song}}, \bibinfo
  {author} {\bibfnamefont {C.}~\bibnamefont {Wang}}, \bibinfo {author}
  {\bibfnamefont {M.}~\bibnamefont {Zhang}}, \bibinfo {author} {\bibfnamefont
  {K.}~\bibnamefont {Hu}}, \bibinfo {author} {\bibfnamefont {Z.}~\bibnamefont
  {Tian}}, \bibinfo {author} {\bibfnamefont {W.}~\bibnamefont {Su}}, \bibinfo
  {author} {\bibfnamefont {P.~K.}\ \bibnamefont {Chu}}, \bibinfo {author}
  {\bibfnamefont {W.}~\bibnamefont {Zhang}},\ and\ \bibinfo {author}
  {\bibfnamefont {Z.}~\bibnamefont {Di}},\ }\bibfield  {title} {\bibinfo
  {title} {A versatile approach to create nanobubbles on arbitrary
  two-dimensional materials for imaging exciton localization},\ }\href@noop {}
  {\bibfield  {journal} {\bibinfo  {journal} {Adv.~Mater.~Interfaces}\ }\textbf
  {\bibinfo {volume} {9}},\ \bibinfo {pages} {2201079} (\bibinfo {year}
  {2022})}\BibitemShut {NoStop}%
\bibitem [{\citenamefont {Gastaldo}\ \emph {et~al.}(2023)\citenamefont
  {Gastaldo}, \citenamefont {Varillas}, \citenamefont {Rodr{\'\i}guez},
  \citenamefont {Velick{\`y}}, \citenamefont {Frank},\ and\ \citenamefont
  {Kalb{\'a}{\v{c}}}}]{gast+23npj2dma}%
  \BibitemOpen
  \bibfield  {author} {\bibinfo {author} {\bibfnamefont {M.}~\bibnamefont
  {Gastaldo}}, \bibinfo {author} {\bibfnamefont {J.}~\bibnamefont {Varillas}},
  \bibinfo {author} {\bibfnamefont {{\'A}.}~\bibnamefont {Rodr{\'\i}guez}},
  \bibinfo {author} {\bibfnamefont {M.}~\bibnamefont {Velick{\`y}}}, \bibinfo
  {author} {\bibfnamefont {O.}~\bibnamefont {Frank}},\ and\ \bibinfo {author}
  {\bibfnamefont {M.}~\bibnamefont {Kalb{\'a}{\v{c}}}},\ }\bibfield  {title}
  {\bibinfo {title} {Tunable strain and bandgap in subcritical-sized \ce{MoS2}
  nanobubbles},\ }\href@noop {} {\bibfield  {journal} {\bibinfo  {journal}
  {npj~{2D}~Mater.~Appl.}\ }\textbf {\bibinfo {volume} {7}},\ \bibinfo {pages}
  {71} (\bibinfo {year} {2023})}\BibitemShut {NoStop}%
\bibitem [{\citenamefont {Steinhoff}\ \emph {et~al.}(2025)\citenamefont
  {Steinhoff}, \citenamefont {Wilksen}, \citenamefont {Solovev}, \citenamefont
  {Schneider},\ and\ \citenamefont {Gies}}]{stei+25prb}%
  \BibitemOpen
  \bibfield  {author} {\bibinfo {author} {\bibfnamefont {A.}~\bibnamefont
  {Steinhoff}}, \bibinfo {author} {\bibfnamefont {S.}~\bibnamefont {Wilksen}},
  \bibinfo {author} {\bibfnamefont {I.}~\bibnamefont {Solovev}}, \bibinfo
  {author} {\bibfnamefont {C.}~\bibnamefont {Schneider}},\ and\ \bibinfo
  {author} {\bibfnamefont {C.}~\bibnamefont {Gies}},\ }\bibfield  {title}
  {\bibinfo {title} {Impact of phonon lifetimes on the single-photon
  indistinguishability in quantum emitters based on two-dimensional
  materials},\ }\href@noop {} {\bibfield  {journal} {\bibinfo  {journal}
  {Phys.~Rev.~B}\ }\textbf {\bibinfo {volume} {111}},\ \bibinfo {pages}
  {195431} (\bibinfo {year} {2025})}\BibitemShut {NoStop}%
\bibitem [{\citenamefont {Carmesin}\ \emph {et~al.}(2019)\citenamefont
  {Carmesin}, \citenamefont {Lorke}, \citenamefont {Florian}, \citenamefont
  {Erben}, \citenamefont {Schulz}, \citenamefont {Wehling},\ and\ \citenamefont
  {Jahnke}}]{carm+19nl}%
  \BibitemOpen
  \bibfield  {author} {\bibinfo {author} {\bibfnamefont {C.}~\bibnamefont
  {Carmesin}}, \bibinfo {author} {\bibfnamefont {M.}~\bibnamefont {Lorke}},
  \bibinfo {author} {\bibfnamefont {M.}~\bibnamefont {Florian}}, \bibinfo
  {author} {\bibfnamefont {D.}~\bibnamefont {Erben}}, \bibinfo {author}
  {\bibfnamefont {A.}~\bibnamefont {Schulz}}, \bibinfo {author} {\bibfnamefont
  {T.~O.}\ \bibnamefont {Wehling}},\ and\ \bibinfo {author} {\bibfnamefont
  {F.}~\bibnamefont {Jahnke}},\ }\bibfield  {title} {\bibinfo {title}
  {Quantum-dot-like states in molybdenum disulfide nanostructures due to the
  interplay of local surface wrinkling, strain, and dielectric confinement},\
  }\href@noop {} {\bibfield  {journal} {\bibinfo  {journal} {Nano~Lett.}\
  }\textbf {\bibinfo {volume} {19}},\ \bibinfo {pages} {3182} (\bibinfo {year}
  {2019})}\BibitemShut {NoStop}%
\bibitem [{\citenamefont {Chen}\ \emph {et~al.}(2025)\citenamefont {Chen},
  \citenamefont {Huang}, \citenamefont {Zhang}, \citenamefont {Chen},\ and\
  \citenamefont {Li}}]{chen+25nl}%
  \BibitemOpen
  \bibfield  {author} {\bibinfo {author} {\bibfnamefont {Y.}~\bibnamefont
  {Chen}}, \bibinfo {author} {\bibfnamefont {R.}~\bibnamefont {Huang}},
  \bibinfo {author} {\bibfnamefont {J.}~\bibnamefont {Zhang}}, \bibinfo
  {author} {\bibfnamefont {J.}~\bibnamefont {Chen}},\ and\ \bibinfo {author}
  {\bibfnamefont {Z.-Y.}\ \bibnamefont {Li}},\ }\bibfield  {title} {\bibinfo
  {title} {Sub-10 nm visualization of trions in ultralow-strained monolayer
  mose2},\ }\href@noop {} {\bibfield  {journal} {\bibinfo  {journal}
  {Nano~Lett.}\ }\textbf {\bibinfo {volume} {25}},\ \bibinfo {pages} {11460}
  (\bibinfo {year} {2025})}\BibitemShut {NoStop}%
\bibitem [{\citenamefont {Velja}\ \emph {et~al.}(2024)\citenamefont {Velja},
  \citenamefont {Krumland},\ and\ \citenamefont {Cocchi}}]{velj+24ns}%
  \BibitemOpen
  \bibfield  {author} {\bibinfo {author} {\bibfnamefont {S.}~\bibnamefont
  {Velja}}, \bibinfo {author} {\bibfnamefont {J.}~\bibnamefont {Krumland}},\
  and\ \bibinfo {author} {\bibfnamefont {C.}~\bibnamefont {Cocchi}},\
  }\bibfield  {title} {\bibinfo {title} {Electronic properties of \ce{MoSe2}
  nanowrinkles},\ }\href@noop {} {\bibfield  {journal} {\bibinfo  {journal}
  {Nanoscale}\ }\textbf {\bibinfo {volume} {16}},\ \bibinfo {pages} {7134}
  (\bibinfo {year} {2024})}\BibitemShut {NoStop}%
\bibitem [{\citenamefont {Krumland}\ \emph {et~al.}(2024)\citenamefont
  {Krumland}, \citenamefont {Velja},\ and\ \citenamefont
  {Cocchi}}]{krumland2024quantum}%
  \BibitemOpen
  \bibfield  {author} {\bibinfo {author} {\bibfnamefont {J.}~\bibnamefont
  {Krumland}}, \bibinfo {author} {\bibfnamefont {S.}~\bibnamefont {Velja}},\
  and\ \bibinfo {author} {\bibfnamefont {C.}~\bibnamefont {Cocchi}},\
  }\bibfield  {title} {\bibinfo {title} {Quantum dots in transition metal
  dichalcogenides induced by atomic-scale deformations},\ }\href@noop {}
  {\bibfield  {journal} {\bibinfo  {journal} {ACS~Photon.}\ }\textbf {\bibinfo
  {volume} {11}},\ \bibinfo {pages} {586} (\bibinfo {year} {2024})}\BibitemShut
  {NoStop}%
\bibitem [{\citenamefont {Hohenberg}\ and\ \citenamefont
  {Kohn}(1964)}]{hohenberg1964inhomogeneous}%
  \BibitemOpen
  \bibfield  {author} {\bibinfo {author} {\bibfnamefont {P.}~\bibnamefont
  {Hohenberg}}\ and\ \bibinfo {author} {\bibfnamefont {W.}~\bibnamefont
  {Kohn}},\ }\bibfield  {title} {\bibinfo {title} {Inhomogeneous electron
  gas},\ }\href@noop {} {\bibfield  {journal} {\bibinfo  {journal}
  {Phys.~Rev.}\ }\textbf {\bibinfo {volume} {136}},\ \bibinfo {pages} {B864}
  (\bibinfo {year} {1964})}\BibitemShut {NoStop}%
\bibitem [{\citenamefont {Kohn}\ and\ \citenamefont
  {Sham}(1965)}]{kohn1965self}%
  \BibitemOpen
  \bibfield  {author} {\bibinfo {author} {\bibfnamefont {W.}~\bibnamefont
  {Kohn}}\ and\ \bibinfo {author} {\bibfnamefont {L.~J.}\ \bibnamefont
  {Sham}},\ }\bibfield  {title} {\bibinfo {title} {Self-consistent equations
  including exchange and correlation effects},\ }\href@noop {} {\bibfield
  {journal} {\bibinfo  {journal} {Phys.~Rev.}\ }\textbf {\bibinfo {volume}
  {140}},\ \bibinfo {pages} {A1133} (\bibinfo {year} {1965})}\BibitemShut
  {NoStop}%
\bibitem [{\citenamefont {Giannozzi}\ \emph {et~al.}(2017)\citenamefont
  {Giannozzi}, \citenamefont {Andreussi}, \citenamefont {Brumme}, \citenamefont
  {Bunau}, \citenamefont {Nardelli}, \citenamefont {Calandra}, \citenamefont
  {Car}, \citenamefont {Cavazzoni}, \citenamefont {Ceresoli}, \citenamefont
  {Cococcioni} \emph {et~al.}}]{giannozzi2017advanced}%
  \BibitemOpen
  \bibfield  {author} {\bibinfo {author} {\bibfnamefont {P.}~\bibnamefont
  {Giannozzi}}, \bibinfo {author} {\bibfnamefont {O.}~\bibnamefont
  {Andreussi}}, \bibinfo {author} {\bibfnamefont {T.}~\bibnamefont {Brumme}},
  \bibinfo {author} {\bibfnamefont {O.}~\bibnamefont {Bunau}}, \bibinfo
  {author} {\bibfnamefont {M.~B.}\ \bibnamefont {Nardelli}}, \bibinfo {author}
  {\bibfnamefont {M.}~\bibnamefont {Calandra}}, \bibinfo {author}
  {\bibfnamefont {R.}~\bibnamefont {Car}}, \bibinfo {author} {\bibfnamefont
  {C.}~\bibnamefont {Cavazzoni}}, \bibinfo {author} {\bibfnamefont
  {D.}~\bibnamefont {Ceresoli}}, \bibinfo {author} {\bibfnamefont
  {M.}~\bibnamefont {Cococcioni}}, \emph {et~al.},\ }\bibfield  {title}
  {\bibinfo {title} {Advanced capabilities for materials modelling with
  {Quantum ESPRESSO}},\ }\href@noop {} {\bibfield  {journal} {\bibinfo
  {journal} {J.~Phys.:~Condens.~Matter.}\ }\textbf {\bibinfo {volume} {29}},\
  \bibinfo {pages} {465901} (\bibinfo {year} {2017})}\BibitemShut {NoStop}%
\bibitem [{\citenamefont {Perdew}\ \emph {et~al.}(1996)\citenamefont {Perdew},
  \citenamefont {Burke},\ and\ \citenamefont {Wang}}]{perdew1996generalized}%
  \BibitemOpen
  \bibfield  {author} {\bibinfo {author} {\bibfnamefont {J.~P.}\ \bibnamefont
  {Perdew}}, \bibinfo {author} {\bibfnamefont {K.}~\bibnamefont {Burke}},\ and\
  \bibinfo {author} {\bibfnamefont {Y.}~\bibnamefont {Wang}},\ }\bibfield
  {title} {\bibinfo {title} {Generalized gradient approximation for the
  exchange-correlation hole of a many-electron system},\ }\href@noop {}
  {\bibfield  {journal} {\bibinfo  {journal} {Phys.~Rev.~B}\ }\textbf {\bibinfo
  {volume} {54}},\ \bibinfo {pages} {16533} (\bibinfo {year}
  {1996})}\BibitemShut {NoStop}%
\bibitem [{\citenamefont {Schlipf}\ and\ \citenamefont
  {Gygi}(2015)}]{schlipf2015optimization}%
  \BibitemOpen
  \bibfield  {author} {\bibinfo {author} {\bibfnamefont {M.}~\bibnamefont
  {Schlipf}}\ and\ \bibinfo {author} {\bibfnamefont {F.}~\bibnamefont {Gygi}},\
  }\bibfield  {title} {\bibinfo {title} {Optimization algorithm for the
  generation of {ONCV} pseudopotentials},\ }\href@noop {} {\bibfield  {journal}
  {\bibinfo  {journal} {Comput.~Phys.~Commun.}\ }\textbf {\bibinfo {volume}
  {196}},\ \bibinfo {pages} {36} (\bibinfo {year} {2015})}\BibitemShut
  {NoStop}%
\bibitem [{\citenamefont {Bennett}\ \emph {et~al.}(2019)\citenamefont
  {Bennett}, \citenamefont {Hudson}, \citenamefont {Metz}, \citenamefont
  {Liang}, \citenamefont {Spurgeon}, \citenamefont {Cui},\ and\ \citenamefont
  {Mason}}]{bennett2019systematic}%
  \BibitemOpen
  \bibfield  {author} {\bibinfo {author} {\bibfnamefont {J.~W.}\ \bibnamefont
  {Bennett}}, \bibinfo {author} {\bibfnamefont {B.~G.}\ \bibnamefont {Hudson}},
  \bibinfo {author} {\bibfnamefont {I.~K.}\ \bibnamefont {Metz}}, \bibinfo
  {author} {\bibfnamefont {D.}~\bibnamefont {Liang}}, \bibinfo {author}
  {\bibfnamefont {S.}~\bibnamefont {Spurgeon}}, \bibinfo {author}
  {\bibfnamefont {Q.}~\bibnamefont {Cui}},\ and\ \bibinfo {author}
  {\bibfnamefont {S.~E.}\ \bibnamefont {Mason}},\ }\bibfield  {title} {\bibinfo
  {title} {A systematic determination of hubbard {U} using the {GBRV} ultrasoft
  pseudopotential set},\ }\href@noop {} {\bibfield  {journal} {\bibinfo
  {journal} {Comp.~Mater.~Sci.}\ }\textbf {\bibinfo {volume} {170}},\ \bibinfo
  {pages} {109137} (\bibinfo {year} {2019})}\BibitemShut {NoStop}%
\bibitem [{\citenamefont {Pizzi}\ \emph {et~al.}(2020)\citenamefont {Pizzi},
  \citenamefont {Vitale}, \citenamefont {Arita}, \citenamefont {Bl{\"u}gel},
  \citenamefont {Freimuth}, \citenamefont {G{\'e}ranton}, \citenamefont
  {Gibertini}, \citenamefont {Gresch}, \citenamefont {Johnson}, \citenamefont
  {Koretsune} \emph {et~al.}}]{pizzi2020wannier90}%
  \BibitemOpen
  \bibfield  {author} {\bibinfo {author} {\bibfnamefont {G.}~\bibnamefont
  {Pizzi}}, \bibinfo {author} {\bibfnamefont {V.}~\bibnamefont {Vitale}},
  \bibinfo {author} {\bibfnamefont {R.}~\bibnamefont {Arita}}, \bibinfo
  {author} {\bibfnamefont {S.}~\bibnamefont {Bl{\"u}gel}}, \bibinfo {author}
  {\bibfnamefont {F.}~\bibnamefont {Freimuth}}, \bibinfo {author}
  {\bibfnamefont {G.}~\bibnamefont {G{\'e}ranton}}, \bibinfo {author}
  {\bibfnamefont {M.}~\bibnamefont {Gibertini}}, \bibinfo {author}
  {\bibfnamefont {D.}~\bibnamefont {Gresch}}, \bibinfo {author} {\bibfnamefont
  {C.}~\bibnamefont {Johnson}}, \bibinfo {author} {\bibfnamefont
  {T.}~\bibnamefont {Koretsune}}, \emph {et~al.},\ }\bibfield  {title}
  {\bibinfo {title} {Wannier90 as a community code: new features and
  applications},\ }\href@noop {} {\bibfield  {journal} {\bibinfo  {journal}
  {J.~Phys.:~Condens.~Matter.}\ }\textbf {\bibinfo {volume} {32}},\ \bibinfo
  {pages} {165902} (\bibinfo {year} {2020})}\BibitemShut {NoStop}%
\bibitem [{\citenamefont {Ramasubramaniam}(2012)}]{rama12prb}%
  \BibitemOpen
  \bibfield  {author} {\bibinfo {author} {\bibfnamefont {A.}~\bibnamefont
  {Ramasubramaniam}},\ }\bibfield  {title} {\bibinfo {title} {Large excitonic
  effects in monolayers of molybdenum and tungsten dichalcogenides},\
  }\href@noop {} {\bibfield  {journal} {\bibinfo  {journal} {Phys.~Rev.~B}\
  }\textbf {\bibinfo {volume} {86}},\ \bibinfo {pages} {115409} (\bibinfo
  {year} {2012})}\BibitemShut {NoStop}%
\bibitem [{\citenamefont {Haastrup}\ \emph {et~al.}(2018)\citenamefont
  {Haastrup}, \citenamefont {Strange}, \citenamefont {Pandey}, \citenamefont
  {Deilmann}, \citenamefont {Schmidt}, \citenamefont {Hinsche}, \citenamefont
  {Gjerding}, \citenamefont {Torelli}, \citenamefont {Larsen}, \citenamefont
  {Riis-Jensen} \emph {et~al.}}]{haastrup2018computational}%
  \BibitemOpen
  \bibfield  {author} {\bibinfo {author} {\bibfnamefont {S.}~\bibnamefont
  {Haastrup}}, \bibinfo {author} {\bibfnamefont {M.}~\bibnamefont {Strange}},
  \bibinfo {author} {\bibfnamefont {M.}~\bibnamefont {Pandey}}, \bibinfo
  {author} {\bibfnamefont {T.}~\bibnamefont {Deilmann}}, \bibinfo {author}
  {\bibfnamefont {P.~S.}\ \bibnamefont {Schmidt}}, \bibinfo {author}
  {\bibfnamefont {N.~F.}\ \bibnamefont {Hinsche}}, \bibinfo {author}
  {\bibfnamefont {M.~N.}\ \bibnamefont {Gjerding}}, \bibinfo {author}
  {\bibfnamefont {D.}~\bibnamefont {Torelli}}, \bibinfo {author} {\bibfnamefont
  {P.~M.}\ \bibnamefont {Larsen}}, \bibinfo {author} {\bibfnamefont {A.~C.}\
  \bibnamefont {Riis-Jensen}}, \emph {et~al.},\ }\bibfield  {title} {\bibinfo
  {title} {The computational {2D} materials database: high-throughput modeling
  and discovery of atomically thin crystals},\ }\href@noop {} {\bibfield
  {journal} {\bibinfo  {journal} {{2D}~Mater.}\ }\textbf {\bibinfo {volume}
  {5}},\ \bibinfo {pages} {042002} (\bibinfo {year} {2018})}\BibitemShut
  {NoStop}%
\bibitem [{\citenamefont {Gjerding}\ \emph {et~al.}(2021)\citenamefont
  {Gjerding}, \citenamefont {Taghizadeh}, \citenamefont {Rasmussen},
  \citenamefont {Ali}, \citenamefont {Bertoldo}, \citenamefont {Deilmann},
  \citenamefont {Kn{\o}sgaard}, \citenamefont {Kruse}, \citenamefont {Larsen},
  \citenamefont {Manti} \emph {et~al.}}]{gjerding2021recent}%
  \BibitemOpen
  \bibfield  {author} {\bibinfo {author} {\bibfnamefont {M.~N.}\ \bibnamefont
  {Gjerding}}, \bibinfo {author} {\bibfnamefont {A.}~\bibnamefont
  {Taghizadeh}}, \bibinfo {author} {\bibfnamefont {A.}~\bibnamefont
  {Rasmussen}}, \bibinfo {author} {\bibfnamefont {S.}~\bibnamefont {Ali}},
  \bibinfo {author} {\bibfnamefont {F.}~\bibnamefont {Bertoldo}}, \bibinfo
  {author} {\bibfnamefont {T.}~\bibnamefont {Deilmann}}, \bibinfo {author}
  {\bibfnamefont {N.~R.}\ \bibnamefont {Kn{\o}sgaard}}, \bibinfo {author}
  {\bibfnamefont {M.}~\bibnamefont {Kruse}}, \bibinfo {author} {\bibfnamefont
  {A.~H.}\ \bibnamefont {Larsen}}, \bibinfo {author} {\bibfnamefont
  {S.}~\bibnamefont {Manti}}, \emph {et~al.},\ }\bibfield  {title} {\bibinfo
  {title} {Recent progress of the computational {2D} materials database
  ({C2DB})},\ }\href@noop {} {\bibfield  {journal} {\bibinfo  {journal}
  {{2D}~Mater.}\ }\textbf {\bibinfo {volume} {8}},\ \bibinfo {pages} {044002}
  (\bibinfo {year} {2021})}\BibitemShut {NoStop}%
\bibitem [{\citenamefont {Bertolazzi}\ \emph {et~al.}(2011)\citenamefont
  {Bertolazzi}, \citenamefont {Brivio},\ and\ \citenamefont
  {Kis}}]{bertolazzi2011stretching}%
  \BibitemOpen
  \bibfield  {author} {\bibinfo {author} {\bibfnamefont {S.}~\bibnamefont
  {Bertolazzi}}, \bibinfo {author} {\bibfnamefont {J.}~\bibnamefont {Brivio}},\
  and\ \bibinfo {author} {\bibfnamefont {A.}~\bibnamefont {Kis}},\ }\bibfield
  {title} {\bibinfo {title} {Stretching and breaking of ultrathin \ce{MoS2}},\
  }\href@noop {} {\bibfield  {journal} {\bibinfo  {journal} {ACS~Nano}\
  }\textbf {\bibinfo {volume} {5}},\ \bibinfo {pages} {9703} (\bibinfo {year}
  {2011})}\BibitemShut {NoStop}%
\bibitem [{\citenamefont {Liu}\ \emph {et~al.}(2014)\citenamefont {Liu},
  \citenamefont {Yan}, \citenamefont {Chen}, \citenamefont {Fan}, \citenamefont
  {Sun}, \citenamefont {Suh}, \citenamefont {Fu}, \citenamefont {Lee},
  \citenamefont {Zhou}, \citenamefont {Tongay} \emph
  {et~al.}}]{liu2014elastic}%
  \BibitemOpen
  \bibfield  {author} {\bibinfo {author} {\bibfnamefont {K.}~\bibnamefont
  {Liu}}, \bibinfo {author} {\bibfnamefont {Q.}~\bibnamefont {Yan}}, \bibinfo
  {author} {\bibfnamefont {M.}~\bibnamefont {Chen}}, \bibinfo {author}
  {\bibfnamefont {W.}~\bibnamefont {Fan}}, \bibinfo {author} {\bibfnamefont
  {Y.}~\bibnamefont {Sun}}, \bibinfo {author} {\bibfnamefont {J.}~\bibnamefont
  {Suh}}, \bibinfo {author} {\bibfnamefont {D.}~\bibnamefont {Fu}}, \bibinfo
  {author} {\bibfnamefont {S.}~\bibnamefont {Lee}}, \bibinfo {author}
  {\bibfnamefont {J.}~\bibnamefont {Zhou}}, \bibinfo {author} {\bibfnamefont
  {S.}~\bibnamefont {Tongay}}, \emph {et~al.},\ }\bibfield  {title} {\bibinfo
  {title} {Elastic properties of chemical-vapor-deposited monolayer \ce{MoS2},
  \ce{WS2}, and their bilayer heterostructures},\ }\href@noop {} {\bibfield
  {journal} {\bibinfo  {journal} {Nano~Lett.}\ }\textbf {\bibinfo {volume}
  {14}},\ \bibinfo {pages} {5097} (\bibinfo {year} {2014})}\BibitemShut
  {NoStop}%
\bibitem [{\citenamefont {Falin}\ \emph {et~al.}(2021)\citenamefont {Falin},
  \citenamefont {Holwill}, \citenamefont {Lv}, \citenamefont {Gan},
  \citenamefont {Cheng}, \citenamefont {Zhang}, \citenamefont {Qian},
  \citenamefont {Barnett}, \citenamefont {Santos}, \citenamefont {Novoselov}
  \emph {et~al.}}]{falin2021mechanical}%
  \BibitemOpen
  \bibfield  {author} {\bibinfo {author} {\bibfnamefont {A.}~\bibnamefont
  {Falin}}, \bibinfo {author} {\bibfnamefont {M.}~\bibnamefont {Holwill}},
  \bibinfo {author} {\bibfnamefont {H.}~\bibnamefont {Lv}}, \bibinfo {author}
  {\bibfnamefont {W.}~\bibnamefont {Gan}}, \bibinfo {author} {\bibfnamefont
  {J.}~\bibnamefont {Cheng}}, \bibinfo {author} {\bibfnamefont
  {R.}~\bibnamefont {Zhang}}, \bibinfo {author} {\bibfnamefont
  {D.}~\bibnamefont {Qian}}, \bibinfo {author} {\bibfnamefont {M.~R.}\
  \bibnamefont {Barnett}}, \bibinfo {author} {\bibfnamefont {E.~J.}\
  \bibnamefont {Santos}}, \bibinfo {author} {\bibfnamefont {K.~S.}\
  \bibnamefont {Novoselov}}, \emph {et~al.},\ }\bibfield  {title} {\bibinfo
  {title} {Mechanical properties of atomically thin tungsten dichalcogenides:
  \ce{WS2}, \ce{WSe2}, and \ce{WTe2}},\ }\href@noop {} {\bibfield  {journal}
  {\bibinfo  {journal} {ACS~Nano}\ }\textbf {\bibinfo {volume} {15}},\ \bibinfo
  {pages} {2600} (\bibinfo {year} {2021})}\BibitemShut {NoStop}%
\bibitem [{\citenamefont {Zeng}\ \emph {et~al.}(2015)\citenamefont {Zeng},
  \citenamefont {Zhang},\ and\ \citenamefont {Tang}}]{zeng2015electronic}%
  \BibitemOpen
  \bibfield  {author} {\bibinfo {author} {\bibfnamefont {F.}~\bibnamefont
  {Zeng}}, \bibinfo {author} {\bibfnamefont {W.-B.}\ \bibnamefont {Zhang}},\
  and\ \bibinfo {author} {\bibfnamefont {B.-Y.}\ \bibnamefont {Tang}},\
  }\bibfield  {title} {\bibinfo {title} {Electronic structures and elastic
  properties of monolayer and bilayer transition metal dichalcogenides {MX}$_2$
  ({M= Mo, W; X= O, S, Se, Te}): a comparative first-principles study},\
  }\href@noop {} {\bibfield  {journal} {\bibinfo  {journal} {Chin.~Phys.~B}\
  }\textbf {\bibinfo {volume} {24}},\ \bibinfo {pages} {097103} (\bibinfo
  {year} {2015})}\BibitemShut {NoStop}%
\bibitem [{\citenamefont {Cooper}\ \emph {et~al.}(2013)\citenamefont {Cooper},
  \citenamefont {Lee}, \citenamefont {Marianetti}, \citenamefont {Wei},
  \citenamefont {Hone},\ and\ \citenamefont {Kysar}}]{cooper2013nonlinear}%
  \BibitemOpen
  \bibfield  {author} {\bibinfo {author} {\bibfnamefont {R.~C.}\ \bibnamefont
  {Cooper}}, \bibinfo {author} {\bibfnamefont {C.}~\bibnamefont {Lee}},
  \bibinfo {author} {\bibfnamefont {C.~A.}\ \bibnamefont {Marianetti}},
  \bibinfo {author} {\bibfnamefont {X.}~\bibnamefont {Wei}}, \bibinfo {author}
  {\bibfnamefont {J.}~\bibnamefont {Hone}},\ and\ \bibinfo {author}
  {\bibfnamefont {J.~W.}\ \bibnamefont {Kysar}},\ }\bibfield  {title} {\bibinfo
  {title} {Nonlinear elastic behavior of two-dimensional molybdenum
  disulfide},\ }\href@noop {} {\bibfield  {journal} {\bibinfo  {journal}
  {Phys.~Rev.~B}\ }\textbf {\bibinfo {volume} {87}},\ \bibinfo {pages} {035423}
  (\bibinfo {year} {2013})}\BibitemShut {NoStop}%
\bibitem [{\citenamefont {Kokalj}(1999)}]{kokalj1999xcrysden}%
  \BibitemOpen
  \bibfield  {author} {\bibinfo {author} {\bibfnamefont {A.}~\bibnamefont
  {Kokalj}},\ }\bibfield  {title} {\bibinfo {title} {{XCrySDen}—a new program
  for displaying crystalline structures and electron densities},\ }\href@noop
  {} {\bibfield  {journal} {\bibinfo  {journal} {J.~Mol.~Graphics~Model.}\
  }\textbf {\bibinfo {volume} {17}},\ \bibinfo {pages} {176} (\bibinfo {year}
  {1999})},\ \bibinfo {note} {code available from
  \url{http://www.xcrysden.org/}.}\BibitemShut {Stop}%
\bibitem [{\citenamefont {O'Neill}(2006)}]{ONeill2006}%
  \BibitemOpen
  \bibfield  {author} {\bibinfo {author} {\bibfnamefont {B.}~\bibnamefont
  {O'Neill}},\ }\href@noop {} {\emph {\bibinfo {title} {Elementary Differential
  Geometry}}}\ (\bibinfo  {publisher} {Academic Press},\ \bibinfo {year}
  {2006})\BibitemShut {NoStop}%
\bibitem [{\citenamefont {Manzeli}\ \emph {et~al.}(2017)\citenamefont
  {Manzeli}, \citenamefont {Ovchinnikov}, \citenamefont {Pasquier},
  \citenamefont {Yazyev},\ and\ \citenamefont {Kis}}]{manzeli20172d}%
  \BibitemOpen
  \bibfield  {author} {\bibinfo {author} {\bibfnamefont {S.}~\bibnamefont
  {Manzeli}}, \bibinfo {author} {\bibfnamefont {D.}~\bibnamefont
  {Ovchinnikov}}, \bibinfo {author} {\bibfnamefont {D.}~\bibnamefont
  {Pasquier}}, \bibinfo {author} {\bibfnamefont {O.~V.}\ \bibnamefont
  {Yazyev}},\ and\ \bibinfo {author} {\bibfnamefont {A.}~\bibnamefont {Kis}},\
  }\bibfield  {title} {\bibinfo {title} {{2D} transition metal
  dichalcogenides},\ }\href@noop {} {\bibfield  {journal} {\bibinfo  {journal}
  {Nat.~Rev.~Mater.}\ }\textbf {\bibinfo {volume} {2}},\ \bibinfo {pages} {1}
  (\bibinfo {year} {2017})}\BibitemShut {NoStop}%
\bibitem [{\citenamefont {Tongay}\ \emph {et~al.}(2013)\citenamefont {Tongay},
  \citenamefont {Suh}, \citenamefont {Ataca}, \citenamefont {Fan},
  \citenamefont {Luce}, \citenamefont {Kang}, \citenamefont {Liu},
  \citenamefont {Ko}, \citenamefont {Raghunathanan}, \citenamefont {Zhou} \emph
  {et~al.}}]{tongay2013defects}%
  \BibitemOpen
  \bibfield  {author} {\bibinfo {author} {\bibfnamefont {S.}~\bibnamefont
  {Tongay}}, \bibinfo {author} {\bibfnamefont {J.}~\bibnamefont {Suh}},
  \bibinfo {author} {\bibfnamefont {C.}~\bibnamefont {Ataca}}, \bibinfo
  {author} {\bibfnamefont {W.}~\bibnamefont {Fan}}, \bibinfo {author}
  {\bibfnamefont {A.}~\bibnamefont {Luce}}, \bibinfo {author} {\bibfnamefont
  {J.~S.}\ \bibnamefont {Kang}}, \bibinfo {author} {\bibfnamefont
  {J.}~\bibnamefont {Liu}}, \bibinfo {author} {\bibfnamefont {C.}~\bibnamefont
  {Ko}}, \bibinfo {author} {\bibfnamefont {R.}~\bibnamefont {Raghunathanan}},
  \bibinfo {author} {\bibfnamefont {J.}~\bibnamefont {Zhou}}, \emph {et~al.},\
  }\bibfield  {title} {\bibinfo {title} {Defects activated photoluminescence in
  two-dimensional semiconductors: interplay between bound, charged and free
  excitons},\ }\href@noop {} {\bibfield  {journal} {\bibinfo  {journal}
  {Sci.~Rep.}\ }\textbf {\bibinfo {volume} {3}},\ \bibinfo {pages} {2657}
  (\bibinfo {year} {2013})}\BibitemShut {NoStop}%
\bibitem [{\citenamefont {Krustok}\ \emph {et~al.}(2017)\citenamefont
  {Krustok}, \citenamefont {Kaupmees}, \citenamefont {Jaaniso}, \citenamefont
  {Kiisk}, \citenamefont {Sildos}, \citenamefont {Li},\ and\ \citenamefont
  {Gong}}]{krustok2017local}%
  \BibitemOpen
  \bibfield  {author} {\bibinfo {author} {\bibfnamefont {J.}~\bibnamefont
  {Krustok}}, \bibinfo {author} {\bibfnamefont {R.}~\bibnamefont {Kaupmees}},
  \bibinfo {author} {\bibfnamefont {R.}~\bibnamefont {Jaaniso}}, \bibinfo
  {author} {\bibfnamefont {V.}~\bibnamefont {Kiisk}}, \bibinfo {author}
  {\bibfnamefont {I.}~\bibnamefont {Sildos}}, \bibinfo {author} {\bibfnamefont
  {B.}~\bibnamefont {Li}},\ and\ \bibinfo {author} {\bibfnamefont
  {Y.}~\bibnamefont {Gong}},\ }\bibfield  {title} {\bibinfo {title} {Local
  strain-induced band gap fluctuations and exciton localization in aged
  \ce{WS2} monolayers},\ }\href@noop {} {\bibfield  {journal} {\bibinfo
  {journal} {AIP~Adv.}\ }\textbf {\bibinfo {volume} {7}} (\bibinfo {year}
  {2017})}\BibitemShut {NoStop}%
\bibitem [{\citenamefont {Ko{\'s}mider}\ \emph {et~al.}(2013)\citenamefont
  {Ko{\'s}mider}, \citenamefont {Gonz{\'a}lez},\ and\ \citenamefont
  {Fern{\'a}ndez-Rossier}}]{kosmider2013large}%
  \BibitemOpen
  \bibfield  {author} {\bibinfo {author} {\bibfnamefont {K.}~\bibnamefont
  {Ko{\'s}mider}}, \bibinfo {author} {\bibfnamefont {J.~W.}\ \bibnamefont
  {Gonz{\'a}lez}},\ and\ \bibinfo {author} {\bibfnamefont {J.}~\bibnamefont
  {Fern{\'a}ndez-Rossier}},\ }\bibfield  {title} {\bibinfo {title} {Large spin
  splitting in the conduction band of transition metal dichalcogenide
  monolayers},\ }\href@noop {} {\bibfield  {journal} {\bibinfo  {journal}
  {Phys.~Rev.~B}\ }\textbf {\bibinfo {volume} {88}},\ \bibinfo {pages} {245436}
  (\bibinfo {year} {2013})}\BibitemShut {NoStop}%
\bibitem [{\citenamefont {Crowley}\ \emph {et~al.}(2016)\citenamefont
  {Crowley}, \citenamefont {Tahir-Kheli},\ and\ \citenamefont
  {Goddard~III}}]{crowley2016resolution}%
  \BibitemOpen
  \bibfield  {author} {\bibinfo {author} {\bibfnamefont {J.~M.}\ \bibnamefont
  {Crowley}}, \bibinfo {author} {\bibfnamefont {J.}~\bibnamefont
  {Tahir-Kheli}},\ and\ \bibinfo {author} {\bibfnamefont {W.~A.}\ \bibnamefont
  {Goddard~III}},\ }\bibfield  {title} {\bibinfo {title} {Resolution of the
  band gap prediction problem for materials design},\ }\href@noop {} {\bibfield
   {journal} {\bibinfo  {journal} {J.~Phys.~Chem.~Lett.}\ }\textbf {\bibinfo
  {volume} {7}},\ \bibinfo {pages} {1198} (\bibinfo {year} {2016})}\BibitemShut
  {NoStop}%
\bibitem [{\citenamefont {Mori-S{\'a}nchez}\ \emph {et~al.}(2008)\citenamefont
  {Mori-S{\'a}nchez}, \citenamefont {Cohen},\ and\ \citenamefont
  {Yang}}]{mori2008localization}%
  \BibitemOpen
  \bibfield  {author} {\bibinfo {author} {\bibfnamefont {P.}~\bibnamefont
  {Mori-S{\'a}nchez}}, \bibinfo {author} {\bibfnamefont {A.~J.}\ \bibnamefont
  {Cohen}},\ and\ \bibinfo {author} {\bibfnamefont {W.}~\bibnamefont {Yang}},\
  }\bibfield  {title} {\bibinfo {title} {Localization and delocalization errors
  in density functional theory and implications for band-gap prediction},\
  }\href@noop {} {\bibfield  {journal} {\bibinfo  {journal} {Phys.~Rev.~Lett.}\
  }\textbf {\bibinfo {volume} {100}},\ \bibinfo {pages} {146401} (\bibinfo
  {year} {2008})}\BibitemShut {NoStop}%
\bibitem [{\citenamefont {Borlido}\ \emph {et~al.}(2020)\citenamefont
  {Borlido}, \citenamefont {Schmidt}, \citenamefont {Huran}, \citenamefont
  {Tran}, \citenamefont {Marques},\ and\ \citenamefont
  {Botti}}]{borlido2020exchange}%
  \BibitemOpen
  \bibfield  {author} {\bibinfo {author} {\bibfnamefont {P.}~\bibnamefont
  {Borlido}}, \bibinfo {author} {\bibfnamefont {J.}~\bibnamefont {Schmidt}},
  \bibinfo {author} {\bibfnamefont {A.~W.}\ \bibnamefont {Huran}}, \bibinfo
  {author} {\bibfnamefont {F.}~\bibnamefont {Tran}}, \bibinfo {author}
  {\bibfnamefont {M.~A.}\ \bibnamefont {Marques}},\ and\ \bibinfo {author}
  {\bibfnamefont {S.}~\bibnamefont {Botti}},\ }\bibfield  {title} {\bibinfo
  {title} {Exchange-correlation functionals for band gaps of solids: benchmark,
  reparametrization and machine learning},\ }\href@noop {} {\bibfield
  {journal} {\bibinfo  {journal} {npj~Comput.~Mater.}\ }\textbf {\bibinfo
  {volume} {6}},\ \bibinfo {pages} {1} (\bibinfo {year} {2020})}\BibitemShut
  {NoStop}%
\bibitem [{\citenamefont {Qiu}\ \emph {et~al.}(2013)\citenamefont {Qiu},
  \citenamefont {Da~Jornada},\ and\ \citenamefont {Louie}}]{qiu2013optical}%
  \BibitemOpen
  \bibfield  {author} {\bibinfo {author} {\bibfnamefont {D.~Y.}\ \bibnamefont
  {Qiu}}, \bibinfo {author} {\bibfnamefont {F.~H.}\ \bibnamefont
  {Da~Jornada}},\ and\ \bibinfo {author} {\bibfnamefont {S.~G.}\ \bibnamefont
  {Louie}},\ }\bibfield  {title} {\bibinfo {title} {Optical spectrum of
  \ce{MoS2}: many-body effects and diversity of exciton states},\ }\href@noop
  {} {\bibfield  {journal} {\bibinfo  {journal} {Phys.~Rev.~Lett.}\ }\textbf
  {\bibinfo {volume} {111}},\ \bibinfo {pages} {216805} (\bibinfo {year}
  {2013})}\BibitemShut {NoStop}%
\bibitem [{\citenamefont {Komsa}\ and\ \citenamefont
  {Krasheninnikov}(2012)}]{komsa2012effects}%
  \BibitemOpen
  \bibfield  {author} {\bibinfo {author} {\bibfnamefont {H.-P.}\ \bibnamefont
  {Komsa}}\ and\ \bibinfo {author} {\bibfnamefont {A.~V.}\ \bibnamefont
  {Krasheninnikov}},\ }\bibfield  {title} {\bibinfo {title} {Effects of
  confinement and environment on the electronic structure and exciton binding
  energy of \ce{MoS2} from first principles},\ }\href@noop {} {\bibfield
  {journal} {\bibinfo  {journal} {Phys.~Rev.~B}\ }\textbf {\bibinfo {volume}
  {86}},\ \bibinfo {pages} {241201} (\bibinfo {year} {2012})}\BibitemShut
  {NoStop}%
\bibitem [{\citenamefont {Reed}\ \emph {et~al.}(1988)\citenamefont {Reed},
  \citenamefont {Randall}, \citenamefont {Aggarwal}, \citenamefont {Matyi},
  \citenamefont {Moore},\ and\ \citenamefont {Wetsel}}]{reed1988observation}%
  \BibitemOpen
  \bibfield  {author} {\bibinfo {author} {\bibfnamefont {M.}~\bibnamefont
  {Reed}}, \bibinfo {author} {\bibfnamefont {J.}~\bibnamefont {Randall}},
  \bibinfo {author} {\bibfnamefont {R.}~\bibnamefont {Aggarwal}}, \bibinfo
  {author} {\bibfnamefont {R.}~\bibnamefont {Matyi}}, \bibinfo {author}
  {\bibfnamefont {T.}~\bibnamefont {Moore}},\ and\ \bibinfo {author}
  {\bibfnamefont {A.}~\bibnamefont {Wetsel}},\ }\bibfield  {title} {\bibinfo
  {title} {Observation of discrete electronic states in a zero-dimensional
  semiconductor nanostructure},\ }\href@noop {} {\bibfield  {journal} {\bibinfo
   {journal} {Phys.~Rev.~Lett.}\ }\textbf {\bibinfo {volume} {60}},\ \bibinfo
  {pages} {535} (\bibinfo {year} {1988})}\BibitemShut {NoStop}%
\bibitem [{\citenamefont {Tonndorf}\ \emph {et~al.}(2015)\citenamefont
  {Tonndorf}, \citenamefont {Schmidt}, \citenamefont {Schneider}, \citenamefont
  {Kern}, \citenamefont {Buscema}, \citenamefont {Steele}, \citenamefont
  {Castellanos-Gomez}, \citenamefont {van~der Zant}, \citenamefont
  {Michaelis~de Vasconcellos},\ and\ \citenamefont
  {Bratschitsch}}]{tonndorf2015single}%
  \BibitemOpen
  \bibfield  {author} {\bibinfo {author} {\bibfnamefont {P.}~\bibnamefont
  {Tonndorf}}, \bibinfo {author} {\bibfnamefont {R.}~\bibnamefont {Schmidt}},
  \bibinfo {author} {\bibfnamefont {R.}~\bibnamefont {Schneider}}, \bibinfo
  {author} {\bibfnamefont {J.}~\bibnamefont {Kern}}, \bibinfo {author}
  {\bibfnamefont {M.}~\bibnamefont {Buscema}}, \bibinfo {author} {\bibfnamefont
  {G.~A.}\ \bibnamefont {Steele}}, \bibinfo {author} {\bibfnamefont
  {A.}~\bibnamefont {Castellanos-Gomez}}, \bibinfo {author} {\bibfnamefont
  {H.~S.}\ \bibnamefont {van~der Zant}}, \bibinfo {author} {\bibfnamefont
  {S.}~\bibnamefont {Michaelis~de Vasconcellos}},\ and\ \bibinfo {author}
  {\bibfnamefont {R.}~\bibnamefont {Bratschitsch}},\ }\bibfield  {title}
  {\bibinfo {title} {Single-photon emission from localized excitons in an
  atomically thin semiconductor},\ }\href@noop {} {\bibfield  {journal}
  {\bibinfo  {journal} {Optica}\ }\textbf {\bibinfo {volume} {2}},\ \bibinfo
  {pages} {347} (\bibinfo {year} {2015})}\BibitemShut {NoStop}%
\bibitem [{\citenamefont {Koperski}\ \emph {et~al.}(2015)\citenamefont
  {Koperski}, \citenamefont {Nogajewski}, \citenamefont {Arora}, \citenamefont
  {Cherkez}, \citenamefont {Mallet}, \citenamefont {Veuillen}, \citenamefont
  {Marcus}, \citenamefont {Kossacki},\ and\ \citenamefont
  {Potemski}}]{koperski2015single}%
  \BibitemOpen
  \bibfield  {author} {\bibinfo {author} {\bibfnamefont {M.}~\bibnamefont
  {Koperski}}, \bibinfo {author} {\bibfnamefont {K.}~\bibnamefont
  {Nogajewski}}, \bibinfo {author} {\bibfnamefont {A.}~\bibnamefont {Arora}},
  \bibinfo {author} {\bibfnamefont {V.}~\bibnamefont {Cherkez}}, \bibinfo
  {author} {\bibfnamefont {P.}~\bibnamefont {Mallet}}, \bibinfo {author}
  {\bibfnamefont {J.-Y.}\ \bibnamefont {Veuillen}}, \bibinfo {author}
  {\bibfnamefont {J.}~\bibnamefont {Marcus}}, \bibinfo {author} {\bibfnamefont
  {P.}~\bibnamefont {Kossacki}},\ and\ \bibinfo {author} {\bibfnamefont
  {M.}~\bibnamefont {Potemski}},\ }\bibfield  {title} {\bibinfo {title} {Single
  photon emitters in exfoliated \ce{WSe2} structures},\ }\href@noop {}
  {\bibfield  {journal} {\bibinfo  {journal} {Nature~Nanotechnol.}\ }\textbf
  {\bibinfo {volume} {10}},\ \bibinfo {pages} {503} (\bibinfo {year}
  {2015})}\BibitemShut {NoStop}%
\bibitem [{\citenamefont {Linhart}\ \emph {et~al.}(2019)\citenamefont
  {Linhart}, \citenamefont {Paur}, \citenamefont {Smejkal}, \citenamefont
  {Burgd{\"o}rfer}, \citenamefont {Mueller},\ and\ \citenamefont
  {Libisch}}]{linhart2019localized}%
  \BibitemOpen
  \bibfield  {author} {\bibinfo {author} {\bibfnamefont {L.}~\bibnamefont
  {Linhart}}, \bibinfo {author} {\bibfnamefont {M.}~\bibnamefont {Paur}},
  \bibinfo {author} {\bibfnamefont {V.}~\bibnamefont {Smejkal}}, \bibinfo
  {author} {\bibfnamefont {J.}~\bibnamefont {Burgd{\"o}rfer}}, \bibinfo
  {author} {\bibfnamefont {T.}~\bibnamefont {Mueller}},\ and\ \bibinfo {author}
  {\bibfnamefont {F.}~\bibnamefont {Libisch}},\ }\bibfield  {title} {\bibinfo
  {title} {Localized intervalley defect excitons as single-photon emitters in
  \ce{WSe2}},\ }\href@noop {} {\bibfield  {journal} {\bibinfo  {journal}
  {Phys.~Rev.~Lett.}\ }\textbf {\bibinfo {volume} {123}},\ \bibinfo {pages}
  {146401} (\bibinfo {year} {2019})}\BibitemShut {NoStop}%
\bibitem [{\citenamefont {Moody}\ \emph {et~al.}(2018)\citenamefont {Moody},
  \citenamefont {Tran}, \citenamefont {Lu}, \citenamefont {Autry},
  \citenamefont {Fraser}, \citenamefont {Mirin}, \citenamefont {Yang},
  \citenamefont {Li},\ and\ \citenamefont {Silverman}}]{moody2018microsecond}%
  \BibitemOpen
  \bibfield  {author} {\bibinfo {author} {\bibfnamefont {G.}~\bibnamefont
  {Moody}}, \bibinfo {author} {\bibfnamefont {K.}~\bibnamefont {Tran}},
  \bibinfo {author} {\bibfnamefont {X.}~\bibnamefont {Lu}}, \bibinfo {author}
  {\bibfnamefont {T.}~\bibnamefont {Autry}}, \bibinfo {author} {\bibfnamefont
  {J.~M.}\ \bibnamefont {Fraser}}, \bibinfo {author} {\bibfnamefont {R.~P.}\
  \bibnamefont {Mirin}}, \bibinfo {author} {\bibfnamefont {L.}~\bibnamefont
  {Yang}}, \bibinfo {author} {\bibfnamefont {X.}~\bibnamefont {Li}},\ and\
  \bibinfo {author} {\bibfnamefont {K.~L.}\ \bibnamefont {Silverman}},\
  }\bibfield  {title} {\bibinfo {title} {Microsecond valley lifetime of
  defect-bound excitons in monolayer \ce{WSe2}},\ }\href@noop {} {\bibfield
  {journal} {\bibinfo  {journal} {Phys.~Rev.~Lett.}\ }\textbf {\bibinfo
  {volume} {121}},\ \bibinfo {pages} {057403} (\bibinfo {year}
  {2018})}\BibitemShut {NoStop}%
\bibitem [{\citenamefont {Srivastava}\ \emph {et~al.}(2015)\citenamefont
  {Srivastava}, \citenamefont {Sidler}, \citenamefont {Allain}, \citenamefont
  {Lembke}, \citenamefont {Kis},\ and\ \citenamefont
  {{\.I}mamo{\u{g}}lu}}]{srivastava2015optically}%
  \BibitemOpen
  \bibfield  {author} {\bibinfo {author} {\bibfnamefont {A.}~\bibnamefont
  {Srivastava}}, \bibinfo {author} {\bibfnamefont {M.}~\bibnamefont {Sidler}},
  \bibinfo {author} {\bibfnamefont {A.~V.}\ \bibnamefont {Allain}}, \bibinfo
  {author} {\bibfnamefont {D.~S.}\ \bibnamefont {Lembke}}, \bibinfo {author}
  {\bibfnamefont {A.}~\bibnamefont {Kis}},\ and\ \bibinfo {author}
  {\bibfnamefont {A.}~\bibnamefont {{\.I}mamo{\u{g}}lu}},\ }\bibfield  {title}
  {\bibinfo {title} {Optically active quantum dots in monolayer \ce{WSe2}},\
  }\href@noop {} {\bibfield  {journal} {\bibinfo  {journal}
  {Nature~Nanotechnol.}\ }\textbf {\bibinfo {volume} {10}},\ \bibinfo {pages}
  {491} (\bibinfo {year} {2015})}\BibitemShut {NoStop}%
\bibitem [{\citenamefont {Chakraborty}\ \emph {et~al.}(2015)\citenamefont
  {Chakraborty}, \citenamefont {Kinnischtzke}, \citenamefont {Goodfellow},
  \citenamefont {Beams},\ and\ \citenamefont
  {Vamivakas}}]{chakraborty2015voltage}%
  \BibitemOpen
  \bibfield  {author} {\bibinfo {author} {\bibfnamefont {C.}~\bibnamefont
  {Chakraborty}}, \bibinfo {author} {\bibfnamefont {L.}~\bibnamefont
  {Kinnischtzke}}, \bibinfo {author} {\bibfnamefont {K.~M.}\ \bibnamefont
  {Goodfellow}}, \bibinfo {author} {\bibfnamefont {R.}~\bibnamefont {Beams}},\
  and\ \bibinfo {author} {\bibfnamefont {A.~N.}\ \bibnamefont {Vamivakas}},\
  }\bibfield  {title} {\bibinfo {title} {Voltage-controlled quantum light from
  an atomically thin semiconductor},\ }\href@noop {} {\bibfield  {journal}
  {\bibinfo  {journal} {Nature~Nanotechnol.}\ }\textbf {\bibinfo {volume}
  {10}},\ \bibinfo {pages} {507} (\bibinfo {year} {2015})}\BibitemShut
  {NoStop}%
\bibitem [{\citenamefont {Krumland}\ and\ \citenamefont
  {Cocchi}(2021)}]{krum-cocc21es}%
  \BibitemOpen
  \bibfield  {author} {\bibinfo {author} {\bibfnamefont {J.}~\bibnamefont
  {Krumland}}\ and\ \bibinfo {author} {\bibfnamefont {C.}~\bibnamefont
  {Cocchi}},\ }\bibfield  {title} {\bibinfo {title} {Conditions for electronic
  hybridization between transition-metal dichalcogenide monolayers and
  physisorbed carbon-conjugated molecules},\ }\href@noop {} {\bibfield
  {journal} {\bibinfo  {journal} {Electron.~Struct.}\ }\textbf {\bibinfo
  {volume} {3}},\ \bibinfo {pages} {044003} (\bibinfo {year}
  {2021})}\BibitemShut {NoStop}%
\bibitem [{\citenamefont {Goswami}\ \emph {et~al.}(2021)\citenamefont
  {Goswami}, \citenamefont {Bhatt}, \citenamefont {Babu}, \citenamefont {Kaur},
  \citenamefont {Ghorai},\ and\ \citenamefont {Ghosh}}]{gosw+21jpcl}%
  \BibitemOpen
  \bibfield  {author} {\bibinfo {author} {\bibfnamefont {T.}~\bibnamefont
  {Goswami}}, \bibinfo {author} {\bibfnamefont {H.}~\bibnamefont {Bhatt}},
  \bibinfo {author} {\bibfnamefont {K.~J.}\ \bibnamefont {Babu}}, \bibinfo
  {author} {\bibfnamefont {G.}~\bibnamefont {Kaur}}, \bibinfo {author}
  {\bibfnamefont {N.}~\bibnamefont {Ghorai}},\ and\ \bibinfo {author}
  {\bibfnamefont {H.~N.}\ \bibnamefont {Ghosh}},\ }\bibfield  {title} {\bibinfo
  {title} {Ultrafast insights into high energy ({C} and {D}) excitons in few
  layer \ce{WS2}},\ }\href@noop {} {\bibfield  {journal} {\bibinfo  {journal}
  {J.~Phys.~Chem.~Lett.}\ }\textbf {\bibinfo {volume} {12}},\ \bibinfo {pages}
  {6526} (\bibinfo {year} {2021})}\BibitemShut {NoStop}%
\bibitem [{\citenamefont {Li}\ \emph {et~al.}(2021)\citenamefont {Li},
  \citenamefont {Wu}, \citenamefont {Liu}, \citenamefont {Xu},\ and\
  \citenamefont {Liu}}]{li+21apl}%
  \BibitemOpen
  \bibfield  {author} {\bibinfo {author} {\bibfnamefont {Y.}~\bibnamefont
  {Li}}, \bibinfo {author} {\bibfnamefont {X.}~\bibnamefont {Wu}}, \bibinfo
  {author} {\bibfnamefont {W.}~\bibnamefont {Liu}}, \bibinfo {author}
  {\bibfnamefont {H.}~\bibnamefont {Xu}},\ and\ \bibinfo {author}
  {\bibfnamefont {X.}~\bibnamefont {Liu}},\ }\bibfield  {title} {\bibinfo
  {title} {Revealing the interrelation between {C}-and {A}-exciton dynamics in
  monolayer \ce{WS2} via transient absorption spectroscopy},\ }\href@noop {}
  {\bibfield  {journal} {\bibinfo  {journal} {Appl.~Phys.~Lett.}\ }\textbf
  {\bibinfo {volume} {119}} (\bibinfo {year} {2021})}\BibitemShut {NoStop}%
\bibitem [{\citenamefont {Luo}\ \emph {et~al.}(2018)\citenamefont {Luo},
  \citenamefont {Shepard}, \citenamefont {Ardelean}, \citenamefont {Rhodes},
  \citenamefont {Kim}, \citenamefont {Barmak}, \citenamefont {Hone},\ and\
  \citenamefont {Strauf}}]{luo+18natn}%
  \BibitemOpen
  \bibfield  {author} {\bibinfo {author} {\bibfnamefont {Y.}~\bibnamefont
  {Luo}}, \bibinfo {author} {\bibfnamefont {G.~D.}\ \bibnamefont {Shepard}},
  \bibinfo {author} {\bibfnamefont {J.~V.}\ \bibnamefont {Ardelean}}, \bibinfo
  {author} {\bibfnamefont {D.~A.}\ \bibnamefont {Rhodes}}, \bibinfo {author}
  {\bibfnamefont {B.}~\bibnamefont {Kim}}, \bibinfo {author} {\bibfnamefont
  {K.}~\bibnamefont {Barmak}}, \bibinfo {author} {\bibfnamefont {J.~C.}\
  \bibnamefont {Hone}},\ and\ \bibinfo {author} {\bibfnamefont
  {S.}~\bibnamefont {Strauf}},\ }\bibfield  {title} {\bibinfo {title}
  {Deterministic coupling of site-controlled quantum emitters in monolayer
  \ce{WSe2} to plasmonic nanocavities},\ }\href@noop {} {\bibfield  {journal}
  {\bibinfo  {journal} {Nature~Nanotechnol.}\ }\textbf {\bibinfo {volume}
  {13}},\ \bibinfo {pages} {1137} (\bibinfo {year} {2018})}\BibitemShut
  {NoStop}%
\bibitem [{\citenamefont {Cianci}\ \emph {et~al.}(2023)\citenamefont {Cianci},
  \citenamefont {Blundo}, \citenamefont {Tuzi}, \citenamefont {Pettinari},
  \citenamefont {Olkowska-Pucko}, \citenamefont {Parmenopoulou}, \citenamefont
  {Peeters}, \citenamefont {Miriametro}, \citenamefont {Taniguchi},
  \citenamefont {Watanabe} \emph {et~al.}}]{cianci2023spatially}%
  \BibitemOpen
  \bibfield  {author} {\bibinfo {author} {\bibfnamefont {S.}~\bibnamefont
  {Cianci}}, \bibinfo {author} {\bibfnamefont {E.}~\bibnamefont {Blundo}},
  \bibinfo {author} {\bibfnamefont {F.}~\bibnamefont {Tuzi}}, \bibinfo {author}
  {\bibfnamefont {G.}~\bibnamefont {Pettinari}}, \bibinfo {author}
  {\bibfnamefont {K.}~\bibnamefont {Olkowska-Pucko}}, \bibinfo {author}
  {\bibfnamefont {E.}~\bibnamefont {Parmenopoulou}}, \bibinfo {author}
  {\bibfnamefont {D.~B.}\ \bibnamefont {Peeters}}, \bibinfo {author}
  {\bibfnamefont {A.}~\bibnamefont {Miriametro}}, \bibinfo {author}
  {\bibfnamefont {T.}~\bibnamefont {Taniguchi}}, \bibinfo {author}
  {\bibfnamefont {K.}~\bibnamefont {Watanabe}}, \emph {et~al.},\ }\bibfield
  {title} {\bibinfo {title} {Spatially controlled single photon emitters in
  {hBN}-capped \ce{WS2} domes},\ }\href@noop {} {\bibfield  {journal} {\bibinfo
   {journal} {Adv.~Opt.~Mater.}\ }\textbf {\bibinfo {volume} {11}},\ \bibinfo
  {pages} {2202953} (\bibinfo {year} {2023})}\BibitemShut {NoStop}%
\bibitem [{\citenamefont {Tedeschi}\ \emph {et~al.}(2019)\citenamefont
  {Tedeschi}, \citenamefont {Blundo}, \citenamefont {Felici}, \citenamefont
  {Pettinari}, \citenamefont {Liu}, \citenamefont {Yildrim}, \citenamefont
  {Petroni}, \citenamefont {Zhang}, \citenamefont {Zhu}, \citenamefont
  {Sennato} \emph {et~al.}}]{tedeschi2019controlled}%
  \BibitemOpen
  \bibfield  {author} {\bibinfo {author} {\bibfnamefont {D.}~\bibnamefont
  {Tedeschi}}, \bibinfo {author} {\bibfnamefont {E.}~\bibnamefont {Blundo}},
  \bibinfo {author} {\bibfnamefont {M.}~\bibnamefont {Felici}}, \bibinfo
  {author} {\bibfnamefont {G.}~\bibnamefont {Pettinari}}, \bibinfo {author}
  {\bibfnamefont {B.}~\bibnamefont {Liu}}, \bibinfo {author} {\bibfnamefont
  {T.}~\bibnamefont {Yildrim}}, \bibinfo {author} {\bibfnamefont
  {E.}~\bibnamefont {Petroni}}, \bibinfo {author} {\bibfnamefont
  {C.}~\bibnamefont {Zhang}}, \bibinfo {author} {\bibfnamefont
  {Y.}~\bibnamefont {Zhu}}, \bibinfo {author} {\bibfnamefont {S.}~\bibnamefont
  {Sennato}}, \emph {et~al.},\ }\bibfield  {title} {\bibinfo {title}
  {Controlled micro/nanodome formation in proton-irradiated bulk
  transition-metal dichalcogenides},\ }\href@noop {} {\bibfield  {journal}
  {\bibinfo  {journal} {Adv.~Mater.}\ }\textbf {\bibinfo {volume} {31}},\
  \bibinfo {pages} {1903795} (\bibinfo {year} {2019})}\BibitemShut {NoStop}%
\bibitem [{\citenamefont {Pucko}\ \emph {et~al.}(2022)\citenamefont {Pucko},
  \citenamefont {Blundo}, \citenamefont {Zawadzka}, \citenamefont {Cianci},
  \citenamefont {Vaclavkova}, \citenamefont {Kapu{\'s}ci{\'n}ski},
  \citenamefont {Jana}, \citenamefont {Pettinari}, \citenamefont {Felici},
  \citenamefont {Nogajewski} \emph {et~al.}}]{pucko2022excitons}%
  \BibitemOpen
  \bibfield  {author} {\bibinfo {author} {\bibfnamefont {K.~O.}\ \bibnamefont
  {Pucko}}, \bibinfo {author} {\bibfnamefont {E.}~\bibnamefont {Blundo}},
  \bibinfo {author} {\bibfnamefont {N.}~\bibnamefont {Zawadzka}}, \bibinfo
  {author} {\bibfnamefont {S.}~\bibnamefont {Cianci}}, \bibinfo {author}
  {\bibfnamefont {D.}~\bibnamefont {Vaclavkova}}, \bibinfo {author}
  {\bibfnamefont {P.}~\bibnamefont {Kapu{\'s}ci{\'n}ski}}, \bibinfo {author}
  {\bibfnamefont {D.}~\bibnamefont {Jana}}, \bibinfo {author} {\bibfnamefont
  {G.}~\bibnamefont {Pettinari}}, \bibinfo {author} {\bibfnamefont
  {M.}~\bibnamefont {Felici}}, \bibinfo {author} {\bibfnamefont
  {K.}~\bibnamefont {Nogajewski}}, \emph {et~al.},\ }\bibfield  {title}
  {\bibinfo {title} {Excitons and trions in {WSSe} monolayers},\ }\href@noop {}
  {\bibfield  {journal} {\bibinfo  {journal} {2D~Mater.}\ }\textbf {\bibinfo
  {volume} {10}},\ \bibinfo {pages} {015018} (\bibinfo {year}
  {2022})}\BibitemShut {NoStop}%
\bibitem [{\citenamefont {Di~Giorgio}\ \emph {et~al.}(2022)\citenamefont
  {Di~Giorgio}, \citenamefont {Blundo}, \citenamefont {Pettinari},
  \citenamefont {Felici}, \citenamefont {Bobba},\ and\ \citenamefont
  {Polimeni}}]{di2022mechanical}%
  \BibitemOpen
  \bibfield  {author} {\bibinfo {author} {\bibfnamefont {C.}~\bibnamefont
  {Di~Giorgio}}, \bibinfo {author} {\bibfnamefont {E.}~\bibnamefont {Blundo}},
  \bibinfo {author} {\bibfnamefont {G.}~\bibnamefont {Pettinari}}, \bibinfo
  {author} {\bibfnamefont {M.}~\bibnamefont {Felici}}, \bibinfo {author}
  {\bibfnamefont {F.}~\bibnamefont {Bobba}},\ and\ \bibinfo {author}
  {\bibfnamefont {A.}~\bibnamefont {Polimeni}},\ }\bibfield  {title} {\bibinfo
  {title} {Mechanical, elastic, and adhesive properties of two-dimensional
  materials: From straining techniques to state-of-the-art local probe
  measurements},\ }\href@noop {} {\bibfield  {journal} {\bibinfo  {journal}
  {Adv.~Mater.~Interfaces}\ }\textbf {\bibinfo {volume} {9}},\ \bibinfo {pages}
  {2102220} (\bibinfo {year} {2022})}\BibitemShut {NoStop}%
\bibitem [{\citenamefont {Cho}\ and\ \citenamefont
  {Berkelbach}(2018)}]{cho2018environmentally}%
  \BibitemOpen
  \bibfield  {author} {\bibinfo {author} {\bibfnamefont {Y.}~\bibnamefont
  {Cho}}\ and\ \bibinfo {author} {\bibfnamefont {T.~C.}\ \bibnamefont
  {Berkelbach}},\ }\bibfield  {title} {\bibinfo {title} {Environmentally
  sensitive theory of electronic and optical transitions in atomically thin
  semiconductors},\ }\href@noop {} {\bibfield  {journal} {\bibinfo  {journal}
  {Phys.~Rev.~B}\ }\textbf {\bibinfo {volume} {97}},\ \bibinfo {pages} {041409}
  (\bibinfo {year} {2018})}\BibitemShut {NoStop}%
\bibitem [{\citenamefont {Raja}\ \emph {et~al.}(2017)\citenamefont {Raja},
  \citenamefont {Chaves}, \citenamefont {Yu}, \citenamefont {Arefe},
  \citenamefont {Hill}, \citenamefont {Rigosi}, \citenamefont {Berkelbach},
  \citenamefont {Nagler}, \citenamefont {Sch{\"u}ller}, \citenamefont {Korn}
  \emph {et~al.}}]{raja2017coulomb}%
  \BibitemOpen
  \bibfield  {author} {\bibinfo {author} {\bibfnamefont {A.}~\bibnamefont
  {Raja}}, \bibinfo {author} {\bibfnamefont {A.}~\bibnamefont {Chaves}},
  \bibinfo {author} {\bibfnamefont {J.}~\bibnamefont {Yu}}, \bibinfo {author}
  {\bibfnamefont {G.}~\bibnamefont {Arefe}}, \bibinfo {author} {\bibfnamefont
  {H.~M.}\ \bibnamefont {Hill}}, \bibinfo {author} {\bibfnamefont {A.~F.}\
  \bibnamefont {Rigosi}}, \bibinfo {author} {\bibfnamefont {T.~C.}\
  \bibnamefont {Berkelbach}}, \bibinfo {author} {\bibfnamefont
  {P.}~\bibnamefont {Nagler}}, \bibinfo {author} {\bibfnamefont
  {C.}~\bibnamefont {Sch{\"u}ller}}, \bibinfo {author} {\bibfnamefont
  {T.}~\bibnamefont {Korn}}, \emph {et~al.},\ }\bibfield  {title} {\bibinfo
  {title} {Coulomb engineering of the bandgap and excitons in two-dimensional
  materials},\ }\href@noop {} {\bibfield  {journal} {\bibinfo  {journal}
  {Nat.~Commun.}\ }\textbf {\bibinfo {volume} {8}},\ \bibinfo {pages} {15251}
  (\bibinfo {year} {2017})}\BibitemShut {NoStop}%
\bibitem [{\citenamefont {Florian}\ \emph {et~al.}(2018)\citenamefont
  {Florian}, \citenamefont {Hartmann}, \citenamefont {Steinhoff}, \citenamefont
  {Klein}, \citenamefont {Holleitner}, \citenamefont {Finley}, \citenamefont
  {Wehling}, \citenamefont {Kaniber},\ and\ \citenamefont
  {Gies}}]{florian2018dielectric}%
  \BibitemOpen
  \bibfield  {author} {\bibinfo {author} {\bibfnamefont {M.}~\bibnamefont
  {Florian}}, \bibinfo {author} {\bibfnamefont {M.}~\bibnamefont {Hartmann}},
  \bibinfo {author} {\bibfnamefont {A.}~\bibnamefont {Steinhoff}}, \bibinfo
  {author} {\bibfnamefont {J.}~\bibnamefont {Klein}}, \bibinfo {author}
  {\bibfnamefont {A.~W.}\ \bibnamefont {Holleitner}}, \bibinfo {author}
  {\bibfnamefont {J.~J.}\ \bibnamefont {Finley}}, \bibinfo {author}
  {\bibfnamefont {T.~O.}\ \bibnamefont {Wehling}}, \bibinfo {author}
  {\bibfnamefont {M.}~\bibnamefont {Kaniber}},\ and\ \bibinfo {author}
  {\bibfnamefont {C.}~\bibnamefont {Gies}},\ }\bibfield  {title} {\bibinfo
  {title} {The dielectric impact of layer distances on exciton and trion
  binding energies in van der {Waals} heterostructures},\ }\href@noop {}
  {\bibfield  {journal} {\bibinfo  {journal} {Nano~Lett.}\ }\textbf {\bibinfo
  {volume} {18}},\ \bibinfo {pages} {2725} (\bibinfo {year}
  {2018})}\BibitemShut {NoStop}%
\bibitem [{\citenamefont {Krumland}\ \emph {et~al.}(2021)\citenamefont
  {Krumland}, \citenamefont {Gil}, \citenamefont {Corni},\ and\ \citenamefont
  {Cocchi}}]{krumland2021layerpcm}%
  \BibitemOpen
  \bibfield  {author} {\bibinfo {author} {\bibfnamefont {J.}~\bibnamefont
  {Krumland}}, \bibinfo {author} {\bibfnamefont {G.}~\bibnamefont {Gil}},
  \bibinfo {author} {\bibfnamefont {S.}~\bibnamefont {Corni}},\ and\ \bibinfo
  {author} {\bibfnamefont {C.}~\bibnamefont {Cocchi}},\ }\bibfield  {title}
  {\bibinfo {title} {{LayerPCM}: An implicit scheme for dielectric screening
  from layered substrates},\ }\href@noop {} {\bibfield  {journal} {\bibinfo
  {journal} {J.~Chem.~Phys.}\ }\textbf {\bibinfo {volume} {154}},\ \bibinfo
  {pages} {224114} (\bibinfo {year} {2021})}\BibitemShut {NoStop}%
\bibitem [{\citenamefont {Tanda~Bonkano}\ \emph {et~al.}(2024)\citenamefont
  {Tanda~Bonkano}, \citenamefont {Palato}, \citenamefont {Krumland},
  \citenamefont {Kovalenko}, \citenamefont {Schwendke}, \citenamefont
  {Guerrini}, \citenamefont {Li}, \citenamefont {Zhu}, \citenamefont {Cocchi},\
  and\ \citenamefont {St{\"a}hler}}]{tanda2024evidence}%
  \BibitemOpen
  \bibfield  {author} {\bibinfo {author} {\bibfnamefont {B.}~\bibnamefont
  {Tanda~Bonkano}}, \bibinfo {author} {\bibfnamefont {S.}~\bibnamefont
  {Palato}}, \bibinfo {author} {\bibfnamefont {J.}~\bibnamefont {Krumland}},
  \bibinfo {author} {\bibfnamefont {S.}~\bibnamefont {Kovalenko}}, \bibinfo
  {author} {\bibfnamefont {P.}~\bibnamefont {Schwendke}}, \bibinfo {author}
  {\bibfnamefont {M.}~\bibnamefont {Guerrini}}, \bibinfo {author}
  {\bibfnamefont {Q.}~\bibnamefont {Li}}, \bibinfo {author} {\bibfnamefont
  {X.}~\bibnamefont {Zhu}}, \bibinfo {author} {\bibfnamefont {C.}~\bibnamefont
  {Cocchi}},\ and\ \bibinfo {author} {\bibfnamefont {J.}~\bibnamefont
  {St{\"a}hler}},\ }\bibfield  {title} {\bibinfo {title} {Evidence for hybrid
  inorganic--organic transitions at the \ce{WS2}/terrylene interface},\
  }\href@noop {} {\bibfield  {journal} {\bibinfo  {journal}
  {Phys.~Status~Solidi~A}\ }\textbf {\bibinfo {volume} {221}},\ \bibinfo
  {pages} {2300346} (\bibinfo {year} {2024})}\BibitemShut {NoStop}%
\bibitem [{\citenamefont {Krumland}\ and\ \citenamefont
  {Cocchi}(2024{\natexlab{a}})}]{krumland2024electronic}%
  \BibitemOpen
  \bibfield  {author} {\bibinfo {author} {\bibfnamefont {J.}~\bibnamefont
  {Krumland}}\ and\ \bibinfo {author} {\bibfnamefont {C.}~\bibnamefont
  {Cocchi}},\ }\bibfield  {title} {\bibinfo {title} {Electronic structure of
  low-dimensional inorganic/organic interfaces: Hybrid density functional
  theory, {G}$_0${W}$_0$, and electrostatic models},\ }\href@noop {} {\bibfield
   {journal} {\bibinfo  {journal} {Phys.~Status~Solidi~A}\ }\textbf {\bibinfo
  {volume} {221}},\ \bibinfo {pages} {2300089} (\bibinfo {year}
  {2024}{\natexlab{a}})}\BibitemShut {NoStop}%
\bibitem [{\citenamefont {Krumland}\ and\ \citenamefont
  {Cocchi}(2024{\natexlab{b}})}]{krumland2024ab}%
  \BibitemOpen
  \bibfield  {author} {\bibinfo {author} {\bibfnamefont {J.}~\bibnamefont
  {Krumland}}\ and\ \bibinfo {author} {\bibfnamefont {C.}~\bibnamefont
  {Cocchi}},\ }\bibfield  {title} {\bibinfo {title} {Ab initio modeling of
  mixed-dimensional heterostructures: A path forward},\ }\href@noop {}
  {\bibfield  {journal} {\bibinfo  {journal} {J.~Phys.~Chem.~Lett.}\ }\textbf
  {\bibinfo {volume} {15}},\ \bibinfo {pages} {5350} (\bibinfo {year}
  {2024}{\natexlab{b}})}\BibitemShut {NoStop}%
\end{thebibliography}

%

\end{document}